\documentclass[letter,11pt]{article}
\usepackage{jheppub} 
\usepackage{physics}
\usepackage{tcolorbox}
\usepackage{tikz}
\usepackage{amsmath}

\title{\boldmath Schwinger-Keldysh effective theory of charge transport: redundancies and systematic $\omega/T$ expansion}

\author[a]{Eren Firat}
\author[a]{Andrew Gomes}
\author[a]{Filippo Nardi}
\author[b]{Riccardo Penco}
\author[a]{Riccardo Rattazzi}
\affiliation[a]{Theoretical Particle Physics Laboratory (LPTP), Institute of Physics, EPFL, Lausanne, Switzerland}
\affiliation[b]{Department of Physics, Carnegie Mellon University
5000 Forbes Ave, Pittsburgh, PA 15213, USA}

\emailAdd{eren.firat@epfl.ch, andrew.gomes@epfl.ch, filippo.nardi@epfl.ch, rpenco@andrew.cmu.edu, riccardo.rattazzi@epfl.ch}

\abstract{We study Schwinger-Keldysh effective field theories (EFTs) for systems with non-Abelian internal symmetries near thermal equilibrium. We consider two approaches that were put forward in the literature---one using a redundant Goldstone parameterization, the other employing an adjoint matter field---and demonstrate their complete equivalence by providing an explicit dictionary and proving their equivalence at the path integral level. Critically, we extend both formalisms to be compatible with the dynamical Kubo-Martin-Schwinger (DKMS) symmetry to all orders in $\hbar \omega /T$, classifying all possible invariant kernels satisfying unitarity constraints. We also establish precise power-counting rules, clarifying the interplay between the semiclassical and hydrodynamic expansions. Our work provides a framework for studying non-Abelian charge transport and fluctuations to arbitrary orders in $\hbar \omega /T$.}

\begin{document}
\maketitle
\flushbottom

\section{Introduction}
\label{sec:intro}

The long-distance and long-time dynamics of systems near thermal equilibrium consist of the transport associated with conserved currents. Indeed, at finite temperature, fluctuations relax very quickly to the bath unless they are protected by a conservation law and thus cannot be destroyed locally. In this case, equilibrium is reached on much longer time-scales via transport phenomena. Hydrodynamics provides the universal Effective Field Theory (EFT) description of such dynamics.

During the last decade, considerable effort has gone into recasting relativistic hydrodynamics in the language of a modern EFT whose effective action is completely determined once the relevant conservation equations are specified. 
This effort was driven in part by advances in nuclear physics~\cite{Jaiswal:2016hex}, cosmology~\cite{Baumann:2010tm}, and astrophysics~\cite{Rezzolla:2013dea} that necessitate a systematic out-of-equilibrium description. It was also motivated by the desire to revisit the foundations of what is arguably the oldest effective field theory in physics in light of our modern understanding. 

The natural framework for tackling out-of-equilibrium problems in quantum field theory is the Schwinger-Keldysh (also known as ``in-in'') formalism~\cite{Bakshi:1962dv,Bakshi:1963bn,PhysRev.126.329,keldysh1965diagram}. This approach requires doubling the number of degrees of freedom and carefully handling the boundary conditions in the far past and future, the former crucially depending on the particular pure or mixed state under consideration. In the EFT approach, reviewed in Section~\ref{sec: SK review}, the aim is to capture all information about these boundary conditions by working with an effective action that, at energy and momentum scales well below the characteristic scales of the state (e.g. temperature $T$, inverse mean free time and path, etc.), admits a local derivative expansion. In the particular case of a thermal state, the derivative expansion within the EFT is organized in powers of $\hbar \omega/T$, $\omega \tau$, and $k \ell$, where $\tau$ and $\ell$ are the respective mean free time and mean free path of microscopic interactions. The first expansion parameter accounts for quantum statistical effects, while the latter two account for classical statistical fluctuations. Generically, $\tau$ is larger than $\hbar/T$, however if the underlying dynamics are strongly coupled these scales are expected to coincide by dimensional analysis.\footnote{While the bound $\tau~\gtrsim~\hbar/T$ appears naturally in the EFT context, it has long been known in the context of many body physics and is called \emph{Planckian bound} in the literature~\cite{Mousatov_2020,Hartnoll:2021ydi}.} 

A thermal state generically breaks a symmetry group of the form $G \times G$ spontaneously down to a diagonal subgroup~\cite{Akyuz:2023lsm}. This symmetry breaking pattern has recently been dubbed {\it strong-to-weak} symmetry breaking~\cite{Lessa:2024gcw,Gu:2024wgc,Huang:2024rml}. The associated Goldstone modes dominate the long-range, late-time dynamics of the system and manifest themselves through the existence of poles at $\omega = k = 0$ in the correlation functions of conserved currents. Another manifestation of finite temperature states is the fact that correlation functions must satisfy the Kubo-Martin-Schwinger (KMS) condition~\cite{doi:10.1143/JPSJ.12.570,PhysRev.115.1342}. At the level of the effective action, this can be ensured by imposing a so-called dynamical KMS (DKMS) symmetry~\cite{Sieberer:2015hba,Glorioso:2017fpd}. 

Implementing the DKMS symmetry in a way that is compatible with the symmetries realized non-linearly on the Goldstone modes is a complex task. In this paper, we will make the simplifying assumption that the coupling between the charge and energy-momentum modes is so weak that we can effectively consider them separately (see \cite{Pavaskar:2021pfo} for such an example where magnons can be decoupled from phonons). We will leave the discussion of the full hydrodynamic theory to future work and focus here on internal symmetries only. One approach put forward in the literature \cite{Kovtun:2014hpa,Haehl:2015pja, Haehl:2015foa,Crossley:2015evo, Glorioso:2017fpd,  Glorioso:2016gph, Jensen:2017kzi, Glorioso:2020loc, Liu:2018kfw, Landry:2019iel} consists of increasing the number of Goldstone modes as if $G \times G$ was completely broken, and at the same time introducing an additional symmetry that is local only in space (in the rest frame of the thermal bath). We review this approach in Section~\ref{sec: redundant parametrization} and enhance previous treatments in the literature by deriving results that are valid at all orders in \(\hbar \omega / T\), shedding new light on the interplay between the additional local symmetry and  DKMS transformations.

In Section~\ref{sec: kernels}, we classify all possible terms in the effective action that are invariant under the DKMS symmetry at all orders in \(\hbar \omega / T\). At small frequencies, it is sufficient to work up to a finite order in \(\hbar \omega / T\) for practical purposes. However, the conceptual advantage of our analysis is that it enables calculations up to an arbitrary order in this expansion.  To carry out these calculations systematically, one must have well-defined power counting rules---an issue we address in Section~\ref{sec: power counting}. This all-order analysis represents the first main result of this paper.

The downside of the approach discussed in the previous paragraph is that it necessitates a departure from the usual way of building EFTs of Goldstone modes. For this reason, a different approach was recently put forward in~\cite{Akyuz:2023lsm,Akyuz:2025bco}, where the additional Goldstone modes are replaced by matter fields transforming in the same representation under the unbroken diagonal symmetries. This novel approach, originally developed only at leading order in $\hbar \omega /T$, does not require the introduction of additional local symmetries and is reminiscent of the way in which, for instance, nucleons are included in chiral perturbation theory~\cite{Weinberg:1996kr}. In Section~\ref{sec: matter approach} we review this second approach, extend it up to arbitrary orders in \(\hbar \omega / T\), and prove its equivalence with the first approach. These results are the second main contribution of this paper.
\\

\noindent {\bf Conventions.} Unless otherwise specified, we work in natural units with $\hbar = k_B = 1$. The adjoint representation transforms under the action of a group element $h$ as $X \to h X h^{-1}$. The Fourier transform in time, and its inverse, are
\begin{equation}
    \mathcal{O}(\omega) = \int dt\, e^{+i\omega t} \mathcal{O}(t) \ , \qquad\qquad  \mathcal{O}(t) = \int \frac{d\omega}{2\pi}\, e^{-i\omega t} \mathcal{O}(\omega) \ .
\end{equation}

\section{Review of the Schwinger-Keldysh Formalism} \label{sec: SK review}

In this section, we will give a very brief overview of the Schwinger-Keldysh formalism and its use in constructing hydrodynamic effective field theories. The EFT for a $U(1)$ fluid and superfluid is studied in \cite{Crossley:2015evo,Glorioso:2017fpd} while the transport of non-Abelian charges using a coset construction is carried out in \cite{Akyuz:2023lsm}. More technical reviews on the construction of the EFT are \cite{Liu:2018kfw,Kovtun:2012rj,Basar:2024srd}.

\subsection{Quantum field theory on the Closed Time Path}

\begin{figure}
    \centering
    \includegraphics[scale=0.3]{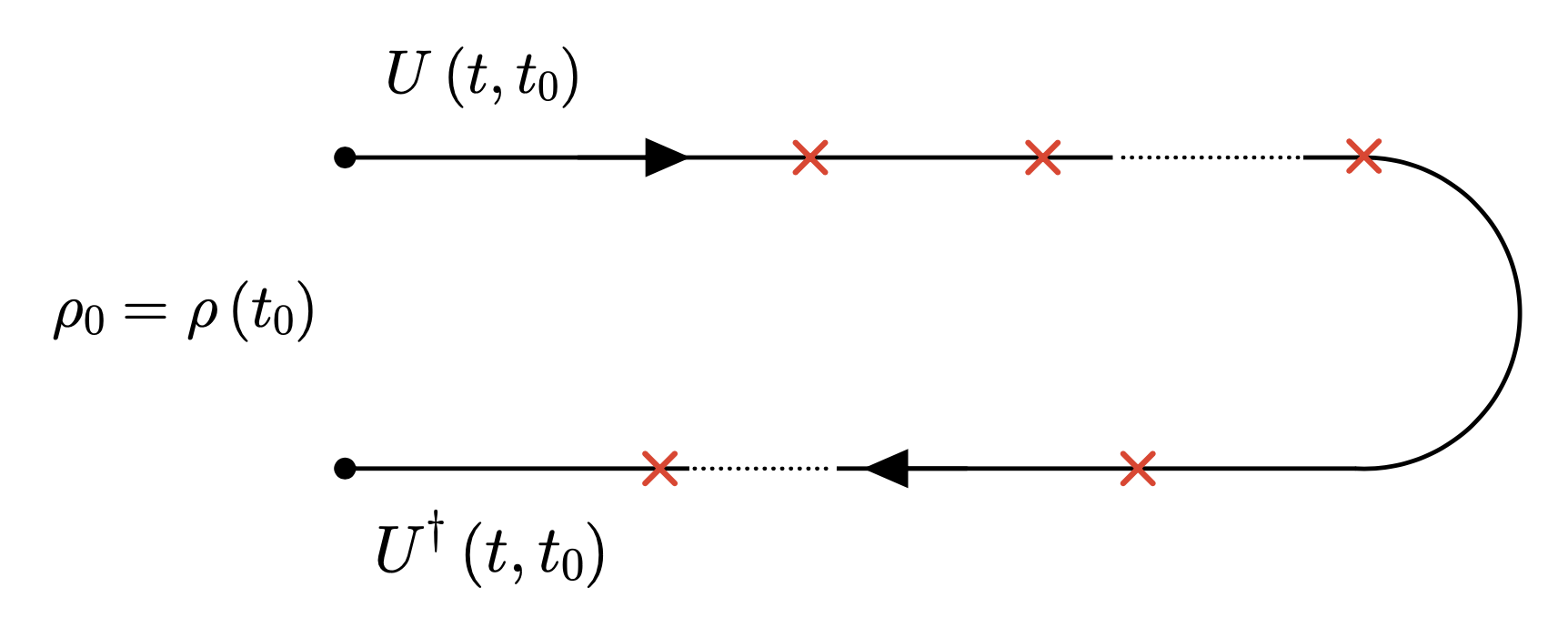}
    \caption{Time contour of the Schwinger-Keldysh path integral. The upper (lower) branch is associated with $1$ ($2$) fields and sources.}
    \label{fig:CTP}
\end{figure}

In the study of non-equilibrium processes, one is generically interested in correlators of the form 
\begin{equation}
\langle \cdots \rangle_{\rho_0} =   \text{Tr}(\cdots\, \rho_0) \ .
\end{equation}
Here $\rho_0$ corresponds to the initial state density matrix, and the dots indicate a generic string of operator insertions, not necessarily time-ordered. A subclass of these correlators is that of the Schwinger-Keldysh type
\begin{equation}
    \text{Tr}(\,\bar{T}(\cdots)\,T(\cdots)\,\rho_0)\ ,
\end{equation}
where the $T$ ($\bar{T}$) symbol indicates (anti-) time-ordering. These correlation functions can be derived from a generating functional of the form
\begin{equation}\label{UVgenfun}
    \mathcal{Z}[J_1,J_2]=\int_{BC}\mathcal{D}\Phi_1 \mathcal{D}\Phi_2\,e^{iS(\Phi_1)-iS(\Phi_2)+i\int J_1\cdot \mathcal{O}_1-i\int J_2.\cdot\mathcal{O}_2} \ .
\end{equation}
This corresponds to a path integral for $\Phi$ (representing all fields of the theory) with a time contour that goes forward and then backward in time as shown in Figure~\ref{fig:CTP}. Fields and sources corresponding to the positively (negatively) oriented time part of the contour are labeled with $1$ ($2$). In this way, taking derivatives with respect to $J_1$ ($J_2$) one can generate an insertion of the corresponding operator in the (anti-) time-ordered branch. The boundary condition $BC$ of the path integral takes into account both the initial state density matrix and the condition that the $1$- and $2$-fields must coincide in the far future, $\Phi_1(t=+\infty)=\Phi_2(t=+\infty)$, as dictated by the trace. In practice, perturbative computations around a thermal state are performed by trading this generating functional for one defined on a single-time contour and without boundary conditions, but with an appropriate $i\varepsilon$ prescription \cite{LeBellac:1996at,Evans:1994gb} (see also Appendix~A of \cite{Akyuz:2023lsm}). Correlation functions with more general time orderings are associated with contours with multiple foldings~\cite{Haehl:2017qfl}. Here we will not discuss this possibility and restrict ourselves to constructing an EFT that can reproduce the Schwinger-Keldysh correlators at low energies. 

As will become clear, a more convenient basis of fields is defined via the so-called Keldysh rotation 
\begin{equation}\label{ra12}
    \Phi_r=\tfrac{1}{2}(\Phi_1+\Phi_2)\ ,\qquad\Phi_a=\Phi_1-\Phi_2\ ,\qquad J_r=\tfrac{1}{2}(J_1+J_2)\ ,\qquad J_a=J_1-J_2\ .
\end{equation}
Unless otherwise specified, for any operator $\mathcal{O}$ we shall use the subscripts $(r, a)$ to refer to the above linear combinations. In particular, the source term becomes
\begin{equation}
    J_1\,\mathcal{O}_1-J_2\,\mathcal{O}_2=J_r\,\mathcal{O}_a+J_a\,\mathcal{O}_r \ ,
\end{equation}
so that taking derivatives with respect to the $r$-source we generate the insertion of an $a$-operator and vice versa. At the level of two-point functions, we have 
\begin{equation}\label{2pts}
\begin{aligned}
    &\langle\mathcal{O}_r(x)\mathcal{O}_r(y) \rangle_{\rho_0} =\frac{1}{2}\langle\{\mathcal{O}(x),\mathcal{O}(y)\} \rangle_{\rho_0} \  &(\text{Keldysh}) \\
    &\langle\mathcal{O}_r(x)\mathcal{O}_a(y) \rangle_{\rho_0} = i\theta(x^0-y^0)\langle[\mathcal{O}(x),\mathcal{O}(y)] \rangle_{\rho_0} \  &(\text{Retarded})\\
    & \langle\mathcal{O}_a(x)\mathcal{O}_r(y) \rangle_{\rho_0} =-i\theta(y^0-x^0)\langle[\mathcal{O}(x),\mathcal{O}(y)] \rangle_{\rho_0} \  \qquad &(\text{Advanced})\\
    &\langle\mathcal{O}_a(x)\mathcal{O}_a(y) \rangle_{\rho_0} = 0 \ .
\end{aligned}
\end{equation}
Note that as a consequence of unitarity, when the latest-time operator is of $a$-type, the correlator vanishes \cite{Liu:2018kfw}. This structural property of the formalism implies that correlators with only $a$-field insertions vanish identically:
\begin{equation}\label{aaa0}
    \langle\mathcal{O}_a^{(1)}(x_1) \cdots \mathcal{O}_a^{(n)}(x_n) \rangle_{\rho_0} = 0.
\end{equation}

\subsection{Kubo-Martin-Schwinger Relations}

Hydrodynamical modes correspond to fluctuations on top of an equilibrium state. Therefore, we will consider as the initial condition the thermal density matrix with inverse temperature $\beta = 1/T$. In an arbitrary frame, the density matrix is
\begin{equation}\label{Thermalrho}
    \rho_0=\frac{e^{-\beta^\mu P_\mu}}{\text{Tr}(e^{-\beta^\mu P_\mu})} \ ,
\end{equation}
where in the fluid rest frame $\beta^\mu = (\beta, \textbf 0)$. This boundary condition, in combination with $CPT$ invariance, translates into the invariance of the generating functional $\mathcal{Z}[J_1', J_2'] = \mathcal{Z}[J_1,J_2]$ under the transformation
\begin{equation}\label{generalKMS}
    J_1'(x^\mu)= \eta\,J_1 \left(-x^\mu +i\frac{\beta^\mu}{2} \right)\ ,\qquad J_2'(x^\mu)=\eta\,J_2\left(-x^\mu - i\frac{\beta^\mu}{2} \right)\ ,
\end{equation}
with $\eta$ the CPT phase of the source.\footnote{While we work exclusively with CPT here, in some cases it may be more convenient to use another discrete symmetry that includes time reversal. Also note that the commutativity of $\rho_0$ with the propagator $e^{-i H t}$ gives an additional freedom in how to define KMS. The form \eqref{generalKMS} is simplest for our purposes.} This property of the generating functional translates into relations among correlation functions. At the level of two-point functions, this is nothing but the Kubo-Martin-Schwinger (KMS) fluctuation-dissipation theorem relating the Fourier transforms of the retarded and Keldysh correlators.
For this reason, \eqref{generalKMS} is called a KMS transformation. Note that this operation maps the $1$- and $2$-currents into themselves while the $r$- and $a$-currents transform in a more complicated way. 

\subsection{The Effective Field Theory Approach}

At distances and time scales much larger than the inverse mean free path and time, one expects the emergence of a universal description for the dynamics of conserved quantities. We use $\phi_r$ and $\phi_a$ to represent the long wavelength/time modes, and to distinguish them from the UV degrees of freedom $\Phi_r$ and $\Phi_a$.  The generating functional can then be written in the form
\begin{equation}\label{effective action ra}
   \mathcal{Z}[J_r,J_a]=\int
   \mathcal{D}\phi_r \mathcal{D}\phi_a
   \, e^{iS_\text{EFT}[\phi_r,\phi_a,J_r,J_a]} \ .
\end{equation}
Note that as commented upon earlier, in the effective theory representation of the generating functional the boundary conditions can been dropped. One enforces the thermality of the initial state by directly asking $\mathcal{Z}$ to be invariant under the KMS transformation (\ref{generalKMS}). This is done in practice by promoting the transformation to act on the dynamical fields and asking for $S_\text{EFT}$ to be invariant. In particular, requiring that the coupling with the external sources be invariant under the transformations of the sources in Eq. \eqref{generalKMS}, we arrive at the following transformation properties for the fields:
\begin{equation}
	 \phi_1'(x) = \eta^{-1} \phi_1 \left(-x^\mu + i \frac{\beta^\mu}{2}\right) \ , \qquad \qquad \phi_2'(x) = \eta^{-1} \phi_2 \left(-x^\mu - i \frac{\beta^\mu}{2}\right) \ .
\end{equation}
This symmetry of the low-energy theory is called Dynamical Kubo-Martin-Schwinger  (DKMS) symmetry in the literature~\cite{Sieberer:2015hba,Glorioso:2017fpd}.

Another structural feature that the EFT must reproduce is the vanishing of the correlator when $a$-type operators are inserted last, and in particular the property \eqref{aaa0}. This property is guaranteed by requiring
\begin{equation}\label{zzzz}
    S_\text{EFT}[\phi_r,\phi_a=0,J_r,J_a=0]=0 \ ,
\end{equation}
along with the condition
\begin{equation}\label{retarded}
    \langle \phi_r(t_1) \phi_{a}(t_2) \rangle_\text{tree} = 0\ , \qquad t_1 > t_2 \ .
\end{equation}
See \cite{Gao:2018bxz} for an all-orders proof.
However, \eqref{retarded} is guaranteed by \eqref{positivity} below together with KMS symmetry, and it is therefore not an independent condition.

Finally, there are two unitarity constraints that the EFT inherits from the microscopic theory. The Schwinger-Keldysh effective action need not be real, and indeed for non-ideal systems (those with dissipation) it is not. Nevertheless, unitarity of the microscopic theory \eqref{UVgenfun} constrains
\begin{equation}\label{unitarity}
    S^*_\text{EFT}[\phi_r,\phi_a,J_r,J_a]=-S_\text{EFT}[\phi_r,-\phi_a,J_r,-J_a] \ .
\end{equation}
In particular this implies that the real part of the effective action is odd in $a$-fields and its imaginary part is even.

Furthermore, the imaginary part of the action should be non-negative \cite{Glorioso:2016gsa}:
\begin{equation}\label{positivity}
    \text{Im}\, S_\text{EFT}[\phi_r,\phi_a,J_r,J_a]\geq 0 \ .
\end{equation}
What this means for the construction of the EFT in practice is that terms of $\mathcal{O}(\phi_a^2)$ should come with a positive sign. Terms higher order in $\phi_a$ are not fixed since, even with a negative sign, \eqref{positivity} can still be maintained due to terms of yet higher order. From this condition one can prove the local version of the second principle of thermodynamics: there exists a current, interpreted as the entropy current, whose divergence is non-negative \cite{Glorioso:2017fpd}.

\section{Symmetry Breaking, Redundancy, and KMS Symmetry} \label{sec: redundant parametrization}

In a zero-temperature QFT with a set of internal symmetries, one can write a low-energy EFT for the gapless degrees of freedom. In addition to the number of gapless modes being fixed by the Goldstone theorem, the form of their interactions is fixed by the pattern of spontaneous symmetry-breaking. This is the usual CCWZ, or coset, construction~\cite{Coleman:1969sm,Callan:1969sn}.

At finite temperatures, the situation becomes more complex. In addition to the gapless modes associated with spontaneously broken symmetries, there are also modes associated with unbroken symmetries. These new modes are responsible for the phenomenon of diffusion. Developing a systematic understanding in terms of a CCWZ-like construction is thus a worthwhile endeavor.

This project, however, comes with difficulties due to the use of the Schwinger-Keldysh formalism, which makes the choice of IR degrees of freedom more obscure. Concretely, there is a direct contradiction between a na\"ive application of the Goldstone theorem and the Schwinger-Keldysh requirement that all fields come in pairs. To understand this, consider the generic symmetry-breaking pattern for a theory invariant under a symmetry group $G$, and broken to a subgroup $H$ in the thermal state. In the Schwinger-Keldysh formalism the symmetry group is doubled so that the pattern of symmetry breaking takes the form~\cite{Akyuz:2023lsm}:
\begin{equation}\label{breaking}
    G_1 \times G_2 \to H_r \ .
\end{equation}
In this context the subscript $r$ denotes diagonal transformations (acting in the same way on $1$- and $2$-fields), while we will use $a$ to denote the transformations which act in opposite ways on $1$- and $2$-fields. Importantly, all $a$-type transformations (also sometimes called `quantum' in the literature) are spontaneously broken when we take the initial density matrix to be a thermal state~\cite{Akyuz:2023lsm}.\footnote{In fact, even if the initial density matrix was a pure state, the final-time boundary condition breaks $G_1 \times G_2 \to G_r$. One indeed finds correlators with the implied poles. However, we are not interested in such a state here.} This pattern was recently dubbed \textit{strong-to-weak} spontaneous symmetry breaking ~\cite{Lessa:2024gcw,Gu:2024wgc,Huang:2024rml}. Depending on the system, some $r$-type (or `classical') transformations may be broken as well.

The coset space associated with \eqref{breaking} can be parametrized by
\begin{equation} \label{eq: coset element}
    \Omega = e^{i \pi_r \cdot  X_r} e^{i \pi_a \cdot X_a} e^{i \phi_a \cdot T_a} \ ,
\end{equation}
where $\cdot$ is the dot product over indices in the adjoint representation, the $X$ are the generators of symmetries broken by the thermal state, the $T$ are the generators of the unbroken group $H$, and finally we define
\begin{equation}\label{gens}
T_r = T_1 + T_2 \ , \qquad T_a = \frac{1}{2} \left(T_1 - T_2 \right) \ .
\end{equation}
Importantly, even though $H$ is unbroken in the thermal state, the $a$-type transformations $H_a$ are broken and thus have associated Goldstone modes $\phi_a$.

As usual the coset transforms under a global transformation on the left as 
\begin{equation}\label{globaltransf}
    \Omega \to g \Omega  h^{-1} \ , \qquad g \in G^L \equiv G_1 \times G_2 \ ,
\end{equation}
where $h(\pi_r, \pi_a, \phi_a, g)$ is a field-dependent element of the unbroken group $H_r$.

\subsection{Introduction of the KMS Partners}

We observe that, due to the breaking pattern \eqref{breaking}, the Goldstones $\phi_a$ do not come with any $r$-field partners. Simply introducing fields $\phi_r$ into the coset would then describe a different symmetry-breaking pattern---one where $H_r$ is trivial---and thus a physically different system. It is also not possible to leave $\phi_a$ alone, arguing that its $r$-partner is outside of the scope of the EFT. One quick argument to dismiss this possibility is simply to observe that an $r$-partner is required in order to calculate the observables associated to the contour in Figure~\ref{fig:CTP}, \textit{e.g.} the two-point functions \eqref{2pts}. Alternatively, one must introduce a partner in order to realize the action of the dynamical KMS transformation on $\phi_a$.

We are thus forced to introduce additional degrees of freedom to the theory beyond the Goldstone bosons na\"ively dictated by the symmetry-breaking pattern.
However, one needs additional assumptions about the low-energy dynamics in order to uniquely define a procedure. In this paper, we will follow the principle of minimality to build the EFT and find that one can take two perspectives, which we shall contrast carefully in Section~\ref{sec: matter approach}.

The first perspective, and perhaps the conceptually simplest one, is to introduce a matter field $\rho_r$ associated with the Goldstone $\phi_a$. By matter field, we mean a field that transforms linearly under $H_r$. 
DKMS transformations commute with $G_1 \times G_2$. This makes it clear that $\rho_r$ should transform in the same representation of $H_r$ as $\phi_a$, namely the adjoint. 
This option, although not the standard approach in the literature, is natural from the perspective of the standard CCWZ construction and was recently explored in \cite{Akyuz:2023lsm}. However, this idea is technically complicated by the difficulty of relating $\phi_a$, which parametrizes a compact manifold, to a matter field, which lives in the Lie algebra, via a DKMS transformation. In Section~\ref{sec: matter approach}, we review and extend this construction.

In this section, we will instead take the second perspective, being the coset analog of the standard construction of Schwinger-Keldysh EFTs and simpler at the technical level~\cite{Crossley:2015evo,Glorioso:2017fpd,Liu:2018kfw}. 
The approach consists in introducing an additional set of compact fields $\phi_r$, supplemented by a local redundancy. The origin and the nature of this redundancy had not been fully understood, although attempts were recently made to uncover its origin~\cite{Huang:2024rml}. However, we will show that its form is dictated entirely by the requirement for compatibility with DKMS symmetry. We further emphasize that, while useful, this redundancy is not essential, which will be made clear in Section~\ref{sec: matter approach}.

The treatment of the `normal' pair of Goldstone modes $\pi_r$ and $\pi_a$ is more straightforward than that of the diffusive mode $\phi_a$, since there is no difficulty in defining their DKMS transformations. To simplify equations for the remainder of this section, we will therefore take $H = G$, so that \eqref{eq: coset element} becomes
\begin{equation} \label{eq: coset element22}
    \Omega = e^{i \phi_a \cdot T_a} \ .
\end{equation}

Let us introduce $\phi_r$ via the redundant coset
\begin{equation}\label{cosetforreal}
    \tilde\Omega = e^{i \phi_a \cdot T_a} e^{i \phi_r \cdot T_r} \equiv e^{i \phi_1 \cdot T_1} e^{i \phi_2 \cdot T_2} \ ,
\end{equation}
supplemented by the local $r$-type right action
\begin{equation}\label{gaugeg}
\tilde\Omega \to \tilde\Omega h^{-1}(x) \ , \qquad h(x) = e^{if(x) \cdot T_r} \in H^R = G^R\ ,
\end{equation}
and $f(x)$ is a local Lie parameter. 
In contrast to \eqref{globaltransf}, this transformation is truly local, in that it is independent of $\phi_a(x)$. 
Note that when $G$ is non-Abelian, $(\phi_1,\phi_2)$ are not the simple linear combinations of $(\phi_r,\phi_a)$ given by \eqref{ra12}. In the $(1,2)$ basis, the transformation acts in the following way:
\begin{equation}\label{rsym}
    \begin{Bmatrix}
    e^{i \phi_1(x) \cdot T} \\
    e^{i \phi_2(x) \cdot T}
  \end{Bmatrix}
  \longrightarrow \begin{Bmatrix}
    e^{i \phi_1(x) \cdot T} e^{i f(x) \cdot T}\\
    e^{i \phi_2(x) \cdot T} e^{i f(x) \cdot T}
  \end{Bmatrix} \ .
\end{equation}

Were we constructing a relativistic field theory, there would be two options for $f(x)$. Either $f$ is a constant, in which case we would end up with a Goldstone mode associated to the unbroken group $G_r$, or it can be any spacetime function, in which case we are back to \eqref{eq: coset element22}. However, due to the rest frame of the thermal bath, there is a third option in which $f$ may be independent of time (in this frame)---maintaining a KMS partner for $\phi_a$, but eliminating the unwelcome propagating Goldstone mode. We show below that due to KMS symmetry, this must in fact be the case.

We can use \eqref{rsym} to define an equivalence class of field configurations related by a local right action, \textit{i.e.}, $G^R$ fibers.
In these redundant variables, the transformation of the fields under DKMS is given by
\begin{equation}\label{eq: KMS transformations}
    \phi_1(x) \to \phi_1 \left(-x^\mu + i \frac{\beta^\mu}{2}\right), \qquad \phi_2(x) \to \phi_2 \left(-x^\mu - i \frac{\beta^\mu}{2}\right) \ .
\end{equation}
However, we must ensure that the action of a DKMS transformation sends one $G^R$ fiber into another in order for it to be well defined. Indeed, consider first transforming under the right action, followed by a DKMS transformation:
\begin{equation}
        \begin{Bmatrix}
    e^{i \phi_1(x) \cdot T} \\
    e^{i \phi_2(x) \cdot T}
  \end{Bmatrix}
  \longrightarrow \begin{Bmatrix}
    e^{i \phi_1(x) \cdot T} e^{i f(x) \cdot T}\\
    e^{i \phi_2(x) \cdot T} e^{i f(x) \cdot T}
  \end{Bmatrix}
  \longrightarrow \begin{Bmatrix}
    e^{i \phi_1\left(-x^\mu + i \frac{\beta^\mu}{2}\right) \cdot T} e^{i f\left(-x^\mu + i \frac{\beta^\mu}{2}\right) \cdot T}\\
    e^{i \phi_2\left(-x^\mu - i \frac{\beta^\mu}{2}\right) \cdot T} e^{i f\left(-x^\mu - i \frac{\beta^\mu}{2}\right) \cdot T}
  \end{Bmatrix} \ .
\end{equation}
This is equivalent to \eqref{eq: KMS transformations} only if,
\begin{equation}\label{restrictgauge}
  f\left(-x^\mu - i \frac{\beta^\mu}{2}\right)=f\left(-x^\mu + i \frac{\beta^\mu}{2}\right) \ .
\end{equation}
Therefore, \eqref{gaugeg} should to be restricted to transformations satisfying \eqref{restrictgauge}. In the rest frame of the bath, this condition amounts to periodicity along the imaginary time direction with a period $\beta$. Furthermore, in the low-frequency regime $\omega \beta \ll 1$, where the effective action is local, only the zero mode of $f$ is accessible and \eqref{restrictgauge} reduces to the requirement that $f$ be time-independent. In the remainder of this paper we shall work in this frame for ease of notation.

\subsection{Invariance under the Right Action}\label{sec: gauge symmetry}

We now have all the ingredients to construct the building blocks of the EFT.
The Maurer-Cartan form of the coset \eqref{cosetforreal} is 
\begin{equation}\label{tilded maurer cartan}
     \tilde\omega \equiv -i \tilde\Omega^{-1} \partial_\mu \tilde\Omega =\tilde D_\mu \phi_1 \cdot T_1 + \tilde D_\mu \phi_2 \cdot T_2 =  \tilde D_\mu \phi_r  \cdot T_r +  \tilde D_\mu \phi_a \cdot T_a \ .
\end{equation}
Here and for the rest of this work, we use the tilde notation to denote expressions in the redundant coset parametrization. Notice that while the relation between $(\phi_1, \phi_2)$ and $(\phi_r, \phi_a)$ is non-linear, the relation between the Goldstone covariant derivatives is (by definition) very simple:
\begin{equation}\label{eq: covariant derivative ar and 12}
    \tilde D_\mu \phi_{r} = \frac{1}{2}(\tilde D_\mu \phi_1 + \tilde D_\mu \phi_2), \qquad \tilde D_\mu \phi_{a} = \tilde D_\mu \phi_1 - \tilde D_\mu \phi_2 \ .
\end{equation}
While the $(r,a)$ basis is the most convenient in which to impose the unitarity constraints as well as implement the redundancy, the DKMS symmetry is more easily expressed in the $(1,2)$ basis.

Under \eqref{gaugeg}, these building blocks transform as
\begin{equation}
    \begin{aligned}
        \tilde D_\mu \phi_r \cdot T_r &\to h \, \tilde D_\mu \phi_r \cdot T_r \, h^{-1} - i h  \partial_\mu  h^{-1} \ , \\
        \tilde D_\mu \phi_a \cdot T_a &\to h \, \tilde D_\mu \phi_a \cdot T_a \, h^{-1} \ .
    \end{aligned}
\end{equation}
Hence, $\tilde D_\mu \phi_a$ as well as $\tilde D_0 \phi_r$ and $\partial_t \tilde D_i\phi_r$ (recall we are taking $\beta^\mu = (\beta, \vec 0)$) transform in the adjoint of $G^R$, while $\tilde D_i \phi_r$ transforms like a connection. We can thus use the latter to form a covariant derivative (written here in the adjoint representation):
\begin{equation}
   \tilde \nabla_i = \partial_i + i \left[ \tilde D_i \phi_r \, , \, \cdot \, \right] \ .
\end{equation}
In summary, the $G^R$ covariant building blocks are
\begin{equation}\label{eq: covariant building block}
    \tilde D_0\phi_r\ , \quad \tilde D_\mu \phi_a \ , \quad  \partial_t\ , \quad \tilde \nabla_i \ , \quad \partial_t \tilde D_i\phi_r\ . 
\end{equation}

It remains to derive how these building blocks transform under the DKMS symmetry. Using \eqref{eq: KMS transformations} and \eqref{eq: covariant derivative ar and 12} we get
\begin{equation} \label{eq: KMS transformations Goldstone approach ar basis}
   \begin{pmatrix} \tilde D_\mu \phi_r'(x) \\ \tilde D_\mu \phi_a'(x) \end{pmatrix} = -\begin{pmatrix} \cosh\left( i \frac{\beta}{2}\partial_t\right) & \frac{1}{2} \sinh\left( i \frac{\beta}{2}\partial_t\right) \\ 2 \sinh\left( i \frac{\beta}{2}\partial_t\right) & \cosh\left( i \frac{\beta}{2}\partial_t \right) \end{pmatrix} \begin{pmatrix} \tilde D_\mu \phi_r(y) \\ \tilde D_\mu \phi_a(y) \end{pmatrix} \Bigg|_{y=-x} \ ,
\end{equation}
It is now just a matter of combining the building blocks in a DKMS invariant way, which is the task we turn to next.

\section{Constructing DKMS-invariant Lagrangians}\label{sec: kernels}

We will now discuss how to combine the above covariant ingredients into an action that is DKMS invariant to arbitrary order in $\omega/T$. To this end, let us begin by considering the general structure of a DKMS invariant term, being a frequency-space convolution of operators with kernel $K$:
\begin{align}\label{kernint}
    \int d^3 \textbf x \int \frac{d^n \omega}{(2\pi)^n} \, 2\pi \delta (\omega_1 + \dots + \omega_n) \, \mathcal{O}^{(1)}_{i_1} (\omega_1, \textbf x) \dots \mathcal{O}^{(n)}_{i_n} (\omega_n, \textbf x) K_{i_1 \dots i_n} (\vec \omega) \ ,
\end{align}
where we have factored out of the kernel a delta function to enforce invariance under time translations. Here, we work with $n$ different operators $\mathcal{O}^{(k)}$, each appearing in two copies labeled with indices $i_k = 1,2$. Each of these operators in real time is built out of a combination of Goldstone covariant derivatives and higher derivatives thereof. For the purposes of the following argument, we will assume that all the indices in \eqref{kernint} are contracted in such a way that the overall term is exactly invariant under the redundancy,\footnote{Since the construction works in an arbitrary number of dimensions, it can also be used to build Wess-Zumino terms, which are invariant only up to a total derivative but can always obtained from an exactly invariant term in one higher dimension~\cite{DHoker:2002aw}.} but the detailed form of such contractions is immaterial for the present question. We will now show that unitarity and KMS invariance impose three constraints on the allowed form of the kernel:
\begin{enumerate}
    \item Due to \eqref{zzzz}, the action should vanish when we identify the two copies of each operator. This translates into a simple sum rule for the kernel:
    \begin{align} \label{eq: kernel constraint 1}
       \sum_{i_1 \dots i_n} K_{i_1 \dots i_n} (\vec \omega) = 0 \ .
    \end{align}
    \item By \eqref{unitarity}, the effective action must satisfy a reality condition that relates configurations with swapped copies: $S[1,2] = - S[2,1]^*$. This imposes the following constraint on the kernel:
    \begin{align} \label{eq: kernel constraint 2}
        K_{i_1 \dots i_n} (\vec \omega) = - (\sigma^1)_{i_1}{}^{j_1} \dots (\sigma^1)_{i_n}{}^{j_n} K_{j_1 \dots j_n}^* ( - \vec \omega) \ ,
    \end{align}
    with $\sigma^1$ the first Pauli matrix.
    \item Finally, the KMS condition in the $(1,2)$ basis leads to:
    \begin{align}\label{eq: kernel constraint 3 non loc}
        M_{i_1}{}^{j_1} (\omega_1) \dots M_{i_n}{}^{j_n} (\omega_n) \, K_{j_1 \dots j_n} (\vec \omega) = K_{i_1 \dots i_n} ( - \vec \omega)\,,
    \end{align}
    where we defined the matrix 
    \begin{equation}
        M_i{}^j(\omega)=\begin{pmatrix}
            e^{-\beta \omega /2} & 0 \\
            0 & e^{+ \beta \omega /2}
        \end{pmatrix} \, .
    \end{equation}
\end{enumerate}

\noindent We can simplify our analysis by rewriting the last condition in a more convenient way. By combining it with the reality condition and performing a contraction with Pauli matrices, we obtain:
\begin{align} \label{eq: kernel constraint 3}
    \tilde M_{i_1}{}^{j_1} (\omega_1) \dots \tilde M_{i_n}{}^{j_n} (\omega_n) \, K_{j_1 \dots j_n} (\vec \omega) = - K_{i_1 \dots i_n}^* (\vec \omega) \ ,
\end{align}
where we have introduced a modified matrix
\begin{equation}
    \tilde M_i{}^j(\omega) = \begin{pmatrix} 0 & e^{+\beta \omega /2} \\ e^{-\beta \omega /2} & 0 \end{pmatrix} \, .
\end{equation}
This reformulation enables the rewriting of the two ``non-local" constraints \eqref{eq: kernel constraint 2} and \eqref{eq: kernel constraint 3 non loc} (relating $\omega$ and $-\omega$) into a local one (and a non-local one).

\subsection{Solving the Constraints}

In order to solve the constraints listed in the previous section, it will be convenient to decompose the kernel into its real and imaginary parts:
\begin{align}
    K_{i_1 \dots i_n} (\vec \omega) = R_{i_1 \dots i_n} (\vec \omega) + i \, I_{i_1 \dots i_n} (\vec \omega)
\end{align}
The constraints \eqref{eq: kernel constraint 1}, \eqref{eq: kernel constraint 2}, and \eqref{eq: kernel constraint 3} imply factorized and separate constraints on $R_{i...}$ and $I_{i...}$. For a kernel with $n$ indices, the most general solution for the real part is a linear combination of $2^{n-1} - 1$ tensors of the form:
\begin{align}\label{real_kern}
	\mathcal{R}_{i_1 \dots i_n} &= \delta_{i_1}^1 \dots \delta_{i_n}^1 - \delta_{i_1}^2 \dots \delta_{i_n}^2 \ , \\
	\mathcal{R}_{i_1 \dots i_k \dots  i_n}^{(k \neq n)} &= n_B(+\omega_n) \delta_{i_1}^1 \dots \delta_{i_{n-1}}^1 \delta_{i_n}^2  + n_B(-\omega_n) \delta_{i_1}^2 \dots \delta_{i_{n-1}}^2 \delta_{i_n}^1  \\
	& \qquad - n_B(+\omega_k) \delta_{i_1}^1 \dots \delta_{i_{k-1}}^1 \delta_{i_k}^2 \delta_{i_{k+1}}^1 \dots \delta_{i_{n}}^1  - n_B(-\omega_k) \delta_{i_1}^2 \dots \delta_{i_{k-1}}^2 \delta_{i_k}^1 \delta_{i_{k+1}}^2 \dots \delta_{i_{n}}^2  \ , \nonumber \\
	\mathcal{R}_{i_1 \dots i_k \dots i_\ell \dots  i_n}^{(k, \ell \neq n)} &= n_B(+\omega_n) \delta_{i_1}^1 \dots \delta_{i_{n-1}}^1 \delta_{i_n}^2  + n_B(-\omega_n) \delta_{i_1}^2 \dots \delta_{i_{n-1}}^2 \delta_{i_n}^1  \\
	&\qquad - n_B(+\omega_k + \omega_\ell)\delta_{i_1}^1 \dots \delta_{i_{k-1}}^1 \delta_{i_k}^2 \delta_{i_{k+1}}^1 \dots \delta_{i_{\ell-1}}^1 \delta_{i_\ell}^2 \delta_{i_{\ell+1}}^1  \dots \delta_{i_{n}}^1  \nonumber \\
	&\qquad \qquad - n_B(-\omega_k - \omega_\ell) \delta_{i_1}^2 \dots \delta_{i_{k-1}}^2 \delta_{i_k}^1 \delta_{i_{k+1}}^2 \dots \delta_{i_{\ell-1}}^2 \delta_{i_\ell}^1 \delta_{i_{\ell+1}}^2  \dots \delta_{i_{n}}^2 \ , \nonumber \\
	  & \vdots \nonumber 
\end{align}
where $n_B( \omega) = (e^{\beta \omega} -1)^{-1}$ is the usual Bose distribution, and we keep going until we have introduced similar tensors that pick out at most $n-2$ of the first $n-1$ indices. By \eqref{eq: kernel constraint 2}, the coefficient that multiplies the tensor $\mathcal{R}_{i_1 \dots i_n}$ must be an even function of $\vec \omega$, while all the other coefficients are odd functions. Notice that these tensors are sparse: they all have 4 non-vanishing components, except for the first one, which has only two.

Similarly, we can express the imaginary part of the kernel as a linear combination of the following tensors:
\begin{align}\label{imag_kern}
	\mathcal{I}_{i_1 \dots i_n} &= \delta_{i_1}^1 \dots \delta_{i_n}^1 + \delta_{i_1}^2 \dots \delta_{i_n}^2 \\
	& \qquad - 2 n_F (+\omega_n) \delta_{i_1}^1 \dots \delta_{i_{n-1}}^1 \delta_{i_n}^2  - 2 n_F(-\omega_n) \delta_{i_1}^2 \dots \delta_{i_{n-1}}^2 \delta_{i_n}^1 \ , \nonumber \\
	\mathcal{I}_{i_1 \dots i_k \dots  i_n}^{(k \neq n)} &=  n_F (+\omega_n) \delta_{i_1}^1 \dots \delta_{i_{n-1}}^1 \delta_{i_n}^2  + n_F(-\omega_n) \delta_{i_1}^2 \dots \delta_{i_{n-1}}^2 \delta_{i_n}^1 \\
	& \qquad - n_F(+\omega_k) \delta_{i_1}^1 \dots \delta_{i_{k-1}}^1 \delta_{i_k}^2 \delta_{i_{k+1}}^1 \dots \delta_{i_{n}}^1  - n_F(-\omega_k) \delta_{i_1}^2 \dots \delta_{i_{k-1}}^2 \delta_{i_k}^1 \delta_{i_{k+1}}^2 \dots \delta_{i_{n}}^2  \ , \nonumber \\
	\mathcal{I}_{i_1 \dots i_k \dots i_\ell \dots  i_n}^{(k, \ell \neq n)} &=  n_F (+\omega_n) \delta_{i_1}^1 \dots \delta_{i_{n-1}}^1 \delta_{i_n}^2  + n_F(-\omega_n) \delta_{i_1}^2 \dots \delta_{i_{n-1}}^2 \delta_{i_n}^1 \\
	&\qquad - n_F(+\omega_k + \omega_\ell)\delta_{i_1}^1 \dots \delta_{i_{k-1}}^1 \delta_{i_k}^2 \delta_{i_{k+1}}^1 \dots \delta_{i_{\ell-1}}^1 \delta_{i_\ell}^2 \delta_{i_{\ell+1}}^1  \dots \delta_{i_{n}}^1  \nonumber \\
	&\qquad \qquad - n_F(-\omega_k - \omega_\ell) \delta_{i_1}^2 \dots \delta_{i_{k-1}}^2 \delta_{i_k}^1 \delta_{i_{k+1}}^2 \dots \delta_{i_{\ell-1}}^2 \delta_{i_\ell}^1 \delta_{i_{\ell+1}}^2  \dots \delta_{i_{n}}^2 \ , \nonumber \\
	  & \vdots \nonumber 
\end{align}
where our tensors now depend on the Fermi distribution $n_F( \omega) = (e^{\beta \omega} + 1)^{-1}$, and once again we keep going until we have introduced similar tensors that pick out at most $n-2$ out of the first $n-1$ indices. This time, all the coefficients multiplying these tensors must be even functions of $\vec \omega$. As we will discuss in the next subsection, in the limit $\omega \ll T$, these kernels can be expanded up to the desired order in the power counting parameter $\omega / T$ and the action becomes local in time at any finite order in this expansion. We have checked that this basis of kernels reproduces at tree level the 2- and 3-point 1PI Green's functions of \cite{Wang:1998wg}.\footnote{To perform the comparison, one should note the technical subtlety that the kernels we computed should be equal to the 1PI Green's functions (in \cite{Wang:1998wg}) up to a change of sign in the $\omega_k$'s.}

In the EFT, it appears more convenient to work in the $(r,a)$ basis, where the power counting (discussed below) of the Lagrangian terms is simpler. However, in general the kernels presented become more complicated  in the $(r,a)$ basis (they typically have more non-zero entries). The two quadratic kernels take a simple form:
\begin{equation}\label{eq: kernels_ra_quad}
    \mathcal{R}=
\begin{pmatrix}
 0 & 1 \\
 1 & 0 
\end{pmatrix}\, , \quad\mathcal{I}=
\begin{pmatrix}
 0 & -\tanh\left( \frac{\beta \omega}{2}\right) \\
 \tanh \left( \frac{\beta \omega}{2}\right) & 1 
\end{pmatrix}\,,
\end{equation}
where we have used $\vec \omega = (\omega, -\omega)$.  $\mathcal R$  and $\mathcal I$ must come with an even coefficient in $\omega$. It is simple to check that they respect the fluctuation-dissipation theorem. In Appendix~\ref{app:kern} we give a basis of kernels at cubic order.

Using the quadratic kernels \eqref{eq: kernels_ra_quad} and the coset ingredients \eqref{eq: covariant building block}, we can obtain an all-orders version of the quadratic action for non-Abelian charge diffusion~\cite{Akyuz:2023lsm}:
\begin{equation}\begin{aligned}\label{eq: quadratic non Abelian eft}
    S &= \int \frac{d\omega}{2\pi} \, d^3 \textbf x \left[ \frac{\chi}{2} \mathcal{R}_{\alpha_1 \alpha_2} \tilde D_0 \phi_{\alpha_1}(\omega)\cdot  \tilde D_0\phi_{\alpha_2}(-\omega) + \frac{i \sigma}{\beta}  \mathcal{I}_{\alpha_1 \alpha_2} \tilde D_i \phi_{\alpha_1}(\omega) \cdot \tilde D_i \phi_{\alpha_2}(-\omega) \right]  \\
    &=  \int \frac{d\omega}{2\pi} d^3 \textbf x \left[ \chi \tilde D_0 \phi_a(\omega) \cdot \tilde D_0\phi_r(-\omega) \right. \\
    &\qquad\left.+ \frac{i  \sigma}{\beta} \left(  -2\tanh \left( \frac{\beta \omega}{2}\right) \tilde D_i\phi_a (\omega) \cdot \tilde D_i\phi_r(-\omega)  + \tilde D_i\phi_a (\omega) \cdot \tilde D_i\phi_a (-\omega)  \right) \right] \ ,
\end{aligned}
\end{equation}
with $\alpha_k \in (r,a)$. Notice that $\tilde D_i \phi_{\alpha_1}(\omega) \cdot \tilde D_i \phi_{\alpha_2}(-\omega)$ is not invariant under the local right action, but becomes so when contracted with $\mathcal{I}_{\alpha_1 \alpha_2}$. To obtain the leading order result in $\omega/T$, we approximate $\tanh(z) \approx z$. In real time this gives the familiar
\begin{equation}\begin{aligned}\label{eq: quadratic non Abelian eft approx}
    S &=  \int dt\, d^3 \textbf x \left[ \chi \tilde D_0 \phi_a  \cdot \tilde D_0\phi_r + \sigma \left(  -\tilde D_i\phi_a \cdot \partial_t \tilde D_i\phi_r  + \frac{i}{\beta} \tilde D_i\phi_a \cdot \tilde D_i\phi_a  \right) \right] \ .
\end{aligned}
\end{equation}
The condition \eqref{positivity} tells us that $\sigma$, the conductivity, is positive. $\chi$ has the interpretation of non-Abelian charge susceptibility.

We make two final comments. First, the procedure outlined above relies on having DKMS pairs $(\mathcal{O}_1 , \mathcal{O}_2)$ or alternatively $(\mathcal{O}_r, \mathcal{O}_a)$. When one considers higher-derivative covariants built out of \eqref{eq: covariant building block}, this is no longer the case. As it stands, one must first form DKMS invariants of different orders in the derivatives and fields, and then match their coefficients with the requirement of $G^R$ invariance. To be more systematic, one should develop a tensor calculus to build higher-derivative, $G^R$-covariant representations of the DKMS symmetry, which can then be combined to form invariants. However, we leave this to future work.

Second, in this section, we demonstrated how to construct actions that are invariant under the DKMS transformations. However, the DKMS transformations contain a shift in imaginary time. They therefore also modify the contour of integration of the path integral. In Appendix~\ref{app: KMS contour}, we discuss this issue in more detail and show that under a few reasonable assumptions, DKMS tranformations are still a symmetry of the path integral.

\subsection{Power Counting Rules}\label{sec: power counting}

Let us use \eqref{eq: quadratic non Abelian eft approx} (so that we are taking $\omega \ll T$) to understand the power counting in the EFT. For simplicity we will take $G=U(1)$, and replace $\chi$ and $\sigma$ with variables that make their dimensionality explicit. Finally, to make contact with other discussions of expansion parameters in the literature, it will be convenient to reintroduce all factors of $\hbar$ explicitly:
\begin{equation}\label{powercount}
    \frac{S}{\hbar} = \frac{p_*}{\hbar} \int dt\, d^3 \textbf x \left[ \frac{k_*^2}{\omega_*} \dot{\phi}_a \dot{\phi}_r - \nabla \phi_a \nabla \dot{\phi}_r + \frac{i}{\beta \hbar} (\nabla \phi_a)^2 \right] \, .
\end{equation}
We have included the momentum scale $p_*$, the frequency scale $\omega_*$, and the wavevector $k_*$. Notice that $k_*$ and $\omega_*$ are defined only up to the ratio $\omega_* / k_*^2$ (being the diffusion coefficient), however we may think of them as being the inverse mean free path and time, respectively. 
As we will see, the factors of $\hbar$ will simply ``go along for the ride,'' and the correct expansion parameters will turn out to be ratios of energy scales, which are the physically relevant quantities, independent of the choice of units.

The final (imaginary) term in Eq. \eqref{powercount} acts as a regulating measure in the Minkowskian path integral. We therefore expect the typical unsuppressed configuration to satisfy:
\begin{equation}
    \frac{\text{Im } S}{\hbar} \sim 1 \, .
\end{equation}
This condition implies the following estimate on the size of a particular mode of $\phi_a$:
\begin{equation}\label{scalea}
    \phi_a \sim \hbar \, \sqrt{\frac{\omega k}{p_* T}} \, ,
\end{equation}
where $\omega$ and $k$ are the mode's respective frequency and wavevector. For $\phi_r$ on the other hand, consider the real part of the action. On a saddle point, $\phi_a$ and thus $S$ vanish identically. However, the path integral will receive contributions from all configurations around the saddle point with $\Delta  \, (\text{Re}\,  S)\sim \hbar$, \textit{i.e.}, those configurations that constructively interfere with one another. Thus, for a typical contributing configuration
\begin{equation}
    \frac{\text{Re } S}{\hbar} \sim 1 \, .
\end{equation}
From this, focusing for instance on the second term in \eqref{powercount} (near on-shell, the first and second terms are the same order) and using the relation \eqref{scalea}, we conclude that\footnote{Properly one should work with the invariant $\dot \phi_r \sim \omega \phi_r$. However, here we aim to make contact with previous discussions in the literature.} 
\begin{equation}\label{scaler}
    \phi_r \sim \sqrt{\frac{k T}{\omega p_*}} \ ,
\end{equation}
Combining \eqref{scalea} and \eqref{scaler}, one finally obtains
\begin{equation}
    \frac{\phi_a}{\phi_r} \sim \frac{\hbar \omega}{T} 
    \, .
\end{equation}
Notice the `$\hbar$' suppression. More physically, $\phi_a$ is small in relation to $\phi_r$ when there is a separation between the time scale of the mode and the characteristic time scale $\hbar / T$. This is also the limit in which the KMS transformations \eqref{eq: KMS transformations Goldstone approach ar basis} becomes local, as all time derivatives come in the combination
\begin{equation}
    \hbar \beta \partial_t \, ,
\end{equation}
and one can Taylor expand \eqref{eq: KMS transformations Goldstone approach ar basis} in this parameter. 
This justifies the simultaneous expansion in powers of time derivatives and $a$-fields that is seen in the literature.

When the microscopic theory has a small parameter, we expect there to be additional expansion parameters in the EFT. For example, this could be a weak coupling or the Boltzmann suppression $e^{-m/T}$ with $m$ the mass gap. In each of these cases, the mean free path/time can be much larger than $\beta$. For instance, in the massless $\lambda |\phi|^4$ theory, one has $\omega_* \sim k_* \sim \lambda^2 T$. Besides the expansion in $\omega/T$ associated with KMS, there will also be expansions in $\omega/\omega_*$ and $k/k_*$. For an interesting discussion of the possible time scales in hydrodynamics, see also~\cite{Mousatov_2020,Hartnoll:2021ydi}.

\section{The Matter Field Approach}\label{sec: matter approach}

In Sections~\ref{sec: redundant parametrization} and \ref{sec: kernels}, we have shown how one can build an EFT for a theory with a non-Abelian symmetry at finite temperature, respecting the KMS symmetry to all orders in $\omega/T$. This was done using a redundant parametrization of the coset, following an idea widespread in the literature (see e.g.~\cite{Liu:2018kfw,Glorioso:2020loc}). In particular, we showed that the redundancy needed to describe cases where the symmetry is not spontaneously broken is constrained. Ensuring compatibility with KMS symmetry implies the redundancy is time-independent. 

It is however interesting to investigate whether there exists a formulation of the EFT that does not rely on a redundant description. Important steps to this end were taken recently in \cite{Akyuz:2023lsm}, with the Goldstone field $\phi_r$ being replaced by a matter field $\rho_r$ that transforms linearly under $G_r$. The main difficulty with this approach---which we will refer to as the ``matter field approach''---is that it is not immediately clear how to formulate a DKMS transformation that mixes the field $\phi_a$ and $\rho_r$, as they belong to different geometric spaces. Specifically, the Goldstone field lives in a compact space, while the matter field does not. In that paper, the authors showed that one can define the leading-order KMS transformations in the high-temperature limit $\omega  \ll T$.  However, it is not immediately clear whether their procedure holds at all orders in $\omega /T$, nor if their approach is fully equivalent to the redundant version that we discussed in Section~\ref{sec: redundant parametrization}.

In this section, we review the approach presented in \cite{Akyuz:2023lsm} and we demonstrate how, using the tools developed in the previous sections, we can derive consistent DKMS transformation rules for the matter field approach to all orders in $\omega/T$. We then provide a dictionary between the two descriptions and prove their equivalence. In light of this, we will comment on the interpretation of the redundancy.

\subsection{Matter Field Approach and Dictionary}\label{sec: dictionary}

As before, we will restrict our discussion to the breaking pattern $G_1 \times G_2 \to G_r$. We seek to build an EFT out of the coset parametrization \eqref{eq: coset element22} and a matter field 
$\rho_r$, which must be in the adjoint representation of $G_r$ for it to serve as a KMS partner of $\phi_a$. This coset should be contrasted with that of \eqref{cosetforreal}, which we denote with $\tilde \Omega$. These two parametrizations are related by
\begin{align}
	\tilde \Omega = \Omega \, e^{i \phi_r \cdot T_r} \ .
\end{align}
We will consistently use tilded and un-tilded variables to distinguish between the variables in each parametrization.

The Maurer-Cartan form in the matter field approach is
\begin{equation}\begin{aligned}\label{omega untilded def}
	\omega= -i \Omega^{-1} \partial_\mu \Omega \equiv  D_\mu \phi_a \cdot T_a + \mathcal{A}_\mu \cdot T_r \ . 
\end{aligned}
\end{equation}
The EFT building blocks that transform covariantly under the $G_r$ are thus 
\begin{equation}\label{matter field coset bbs}
    D_\mu \phi_a,\,  \qquad \nabla_\mu = \partial_\mu + i \left[\mathcal{A_\mu} \, , \, \cdot \, \right] \, , \, \qquad \rho_r\, .
\end{equation}

To draw an equivalence with the redundant parameterization, consider the tilded Maurer-Cartan form \eqref{tilded maurer cartan} written in terms of the un-tilded variables.%
\begin{equation}\begin{aligned}\label{eq: relation between covariant derivatives in two approaches}
\tilde{\omega} &\equiv  \tilde{D}_{\mu} \phi_a \cdot T_a + \tilde{D}_{\mu} \phi_r \cdot T_r  \\
&=  e^{-i \phi_r \cdot T_r} \Big[D_{\mu} \phi_a \cdot T_a  + D_{\mu} \phi_r \cdot T_r  \Big] e^{i \phi_r \cdot T_r} \, , 
\end{aligned}
\end{equation}
where we have defined
\begin{align}
	D_\mu \phi_r \cdot T_r \equiv i e^{i \phi_r \cdot T_r} \partial_{\mu} e^{-i \phi_r \cdot T_r} + \mathcal{A}_{\mu} \cdot T_r \ .\label{eq: covariant phi r no tilde}
\end{align}
We should identify $\rho_r$ with an $r$-type operator that is a singlet under spatial rotations and in the adjoint of $G_r$. The obvious choice is
\begin{equation}\label{identity}
    \rho_r \equiv D_0 \phi_r \ .
\end{equation}

The first two building blocks of \eqref{eq: covariant building block} can thus be written as
\begin{equation}\label{eq: tilde in un-tilde basis}
    \begin{split}
        \tilde D_0 \phi_r \cdot T_r &=  e^{-i \phi_r \cdot T_r} \rho_r\cdot T_r e^{i \phi_r \cdot T_r} \ , \\     
        \tilde D_\mu \phi_a \cdot T_a &=  e^{-i \phi_r \cdot T_r} D_\mu \phi_a\cdot T_a e^{i \phi_r \cdot T_r} \ .
    \end{split}
\end{equation}
To understand why the exponentials $e^{\pm i \phi_r(x) \cdot T_r}$ do not present a problem for translating from the tilded to un-tilded variables, consider the following. The effective action is constructed entirely out of objects (and their covariant derivatives) in the adjoint of $G^R$. All $G^R$ singlets are formed by tracing over products of these objects. If, as in \eqref{eq: tilde in un-tilde basis}, all objects take the form $e^{-i \phi_r \cdot T_r} X \cdot T e^{i \phi_r \cdot T_r}$ where $X$ is an operator transforming in the adjoint of $G_r$, then the exponents will cancel to leave a singlet written entirely in the matter field parameterization.\footnote{We use $T$ to represent generators in the adjoint representation of $G_r$. Both $T_r$ and $T_a$ are in the adjoint so long as $G$ is compact~\cite{Weinberg:1996kr}.}
For example, the term $\tilde D_0 \phi_a \cdot \tilde D_0 \phi_r$ from \eqref{eq: quadratic non Abelian eft approx} becomes $D_0 \phi_a \cdot \rho_r$. Thus the first two entries of our dictionary are
\begin{equation}
    \begin{split}
        \tilde D_0 \phi_r &\longleftrightarrow \rho_r \ , \\
        \tilde D_\mu \phi_a &\longleftrightarrow D_\mu \phi_a \ .
    \end{split}
\end{equation}

To find the dictionary entries for the covariant derivatives $\partial_t$ and  $\tilde \nabla_i$
in \eqref{eq: covariant building block}, consider their action on operators of the form $e^{-i \phi_r \cdot T_r} X\cdot T \, e^{i \phi_r \cdot T_r}$. One finds
\begin{equation}\label{eq: derivatives in tilde and un-tilde}
    \begin{split}
    \partial_t  \left( e^{-i \phi_r \cdot T_r}  X \cdot T \, e^{i \phi_r \cdot T_r} \right) &=  e^{-i \phi_r \cdot T_r} \Bigl( \nabla_t X \cdot T - i \left[ \rho_r \cdot T_r \, , X \cdot T \right] \Bigr)\, e^{i \phi_r \cdot T_r} \ , \\
    \tilde \nabla_i \left( e^{-i \phi_r \cdot T_r} X \cdot T \, e^{i \phi_r \cdot T_r}\right) &= e^{-i \phi_r \cdot T_r}\, \nabla_i X \cdot T \,e^{i \phi_r \cdot T_r}  \ ,
    \end{split}
\end{equation}
where we have used \eqref{identity}. The first equation motivates the definition
\begin{align}
	\mathcal{D}_t = \nabla_t - i \left[ \, \rho_r \, , \, \cdot\,  \right]  = \partial_t  + i \left[ \,\mathcal{A}_0 - \rho_r \, , \, \cdot\,  \right] \,,
\end{align}
which will be useful later. For the final building block of \eqref{eq: covariant building block}, we get
\begin{equation}\label{eq: tilde not tilde dictionary}
    \begin{split}
     \partial_t  \tilde D_i \phi_r \cdot T_r &=  e^{-i \phi_r \cdot T_r}\, \mathcal{D}_t D_i \phi_r \cdot T_r \, e^{i \phi_r \cdot T_r} \\
      &=e^{-i \phi_r \cdot T_r} \left( \mathcal{F}_{0 i}  + \nabla_{i} \rho_r  \right) \cdot T_r\, e^{i \phi_r \cdot T_r} \ ,
    \end{split}
\end{equation}
where $\mathcal{F}_{\mu\nu}$ is the non-Abelian field strength associated to~$\mathcal{A}_\mu$ and in the last line we used the identity
\begin{align}
	\nabla_t D_{i} \phi_r - i [D_0 \phi_r , D_i \phi_r] = \mathcal{F}_{0 i} + \nabla_{i} D_0 \phi_r \ . \label{eq: relation commutator covariant derivative}
\end{align}
We summarize the results \eqref{eq: tilde in un-tilde basis}, \eqref{eq: derivatives in tilde and un-tilde}, and \eqref{eq: tilde not tilde dictionary} in Table~\ref{tab:dictionary}. This dictionary can be used to translate any effective action between the redundant and matter field descriptions.

\begin{table}[h]
    \centering
    \begin{tabular}{|c|c|}
    \hline
    Redundant Approach & Matter Field Approach \\ \hline  \hline
   $\tilde D_\mu \phi_a$ & $D_\mu \phi_a$ \\ \hline
    $\tilde D_0 \phi_r$ & $\rho_r$ \\ \hline
    $\partial_t$ & $\mathcal{D}_t$
    \\ \hline
    $\tilde \nabla_i$ & $\nabla_i$ \\ \hline
    $\partial_t  \tilde D_i \phi_r$ & $\mathcal{F}_{0 i}  + \nabla_{i} \rho_r$ \\ \hline
    \end{tabular}
    \caption{The dictionary between the building blocks \eqref{eq: covariant building block} in the redundant approach (using $\phi_r$) and in the matter field approach (using $\rho_r$). An object $\tilde X$ on the left and the corresponding object $X$ on the right are related by $\tilde X \cdot T = e^{-i \phi_r \cdot T_r} X\cdot T e^{i \phi_r \cdot T_r}$, where $T$ transforms in the adjoint of $G_r$. The covariant derivatives used are $\tilde \nabla_i = \partial_i + i [ \tilde D_i \phi_r \, , \, \cdot \, ] \,$, $\mathcal{D}_t = \nabla_t - i \left[ \, \rho_r \, , \, \cdot\,  \right]\,$, and $\nabla_\mu = \partial_\mu + i \left[\mathcal{A_\mu} \, , \, \cdot \, \right]\,$.
    }
    \label{tab:dictionary}
\end{table}

\subsection{Equivalence of the Path Integral}\label{sec: EOM equivalence}

In this section we explicitly perform the change of variables from the redundant set $(\phi_r,\phi_a)$ to $(\rho_r, \phi_a)$ in the path integral. The measure of the former is given by the Haar measure over the group. This can be constructed starting from the Maurer-Cartan form, which we can write as
\begin{equation}
    -i \tilde{\Omega}^{-1}d\tilde{\Omega}= R^{IJ}(\phi_r) \Bigg[ \bigg(f^{IK}(\phi_a) T_a^ J + h^{IK}(\phi_a)T_r^J\bigg) d\phi_a^K+g^{IK}(\phi_r)T_r^J d\phi_r^K \Bigg],
\end{equation}
with all indices in the adjoint, $f, g, h$ some functions, and $R$ a group rotation with parameters $\phi_r$. The measure of $\phi_r$ and $\phi_a$ factorizes and takes the form,
\begin{equation}
    \mathcal{D}\mu(\phi_r)\mathcal{D}\mu(\phi_a) =\det(g(\phi_r))\prod_{t,\mathbf{x}}d\phi_r(t,\mathbf{x})\det(f(\phi_a))\prod_{t,\mathbf{x}}d\phi_a(t,\mathbf{x}) \ ,
\end{equation}
where $\mu(\cdot)$ denotes the Haar measure. To completely define the measure we need to fix the gauge, removing the right action redundancy. One possible choice consists in fixing the value of $\phi_r$ at some time which we can chose to be $t=0$:
\begin{equation}
    \phi_r^0(\textbf x) \equiv \phi_r(0,\textbf x) = \alpha(\textbf x) \ ,
\end{equation}
whereby the path integral measure is
\begin{equation}\label{measure}
    \int \mathcal{D}\mu(\phi_r) \,\mathcal{D}\mu(\phi_a) \,  \frac{1}{\det g(\phi_r^0)} \delta(\phi_r^0 - \alpha) = \int \mathcal{D}\mu(\phi_r)\bigg|_{t\neq 0} \, \mathcal{D}\mu(\phi_a) \ .
\end{equation}
Notice the delta function has been properly normalized. We first make an intermediate change of variables from $\phi_r$ to $\phi_r^0$ and $\dot \phi_r$, given by
\begin{equation}
    \phi_r(t,\textbf x) = \phi_r^0(\textbf x) + \int_0^t   \dot \phi_r(t', \textbf x) \, dt' \ .
\end{equation}
From this we compute
\begin{equation}
    \frac{\delta \phi_r(t,\textbf x)}{\delta \phi_r^0(\textbf x')} = \delta^3(\textbf x - \textbf x') \, \qquad \frac{\delta \phi_r(t,\textbf x)}{\delta \dot \phi_r(t', \textbf x')} =  \big(\theta(t-t')- \theta(-t')\big) \delta^3(\textbf x - \textbf x') \ .
\end{equation}
The triangular structure of the latter quantity ensures that the Jacobian is one. We now consider a second change of variables where we trade $\dot{\phi}_r$ for $\rho_r$ using \eqref{identity}. This can also be written as 
\begin{equation}
    \rho_r^I = g^{I J}[\phi_r(\phi_r^0,\dot \phi_r)] \dot \phi_r^J + \mathcal{A}_0^I(\phi_a) \ ,
\end{equation}
from which
\begin{equation}
    \frac{\delta \rho^I_r(t,\textbf x)}{\delta \dot \phi^K_r(t', \textbf x')} = \delta^4( x - x') g^{I K}[\phi_r(\phi_r^0,\dot \phi_r)] + \dot \phi_r^J \frac{\partial g^{IJ}}{\partial \phi_r^L} \frac{\delta \phi_r^L}{\delta \dot \phi_r^K}  \ .
\end{equation}
For the same reason as above, the second term does not contribute and the Jacobian is $\det g$. Putting everything together, we find that \eqref{measure} can be written as
\begin{equation}
    \int \mathcal{D}\rho_r \, \mathcal{D}\mu(\phi_a) \ .
\end{equation}

\subsection{Saddle Points}

As a useful illustration of the path integral equivalence, we verify a one-to-one correspondence between their saddle points. We will take the Abelian case here, leaving the conceptually identical but technically more involved non-Abelian proof to Appendix~\ref{app: non Abelian EOM}. 

Consider the equations of motion. Because the action is the same in both descriptions, the equation of motion obtained from the variation with respect to $\phi_a$ is the same. On the other hand, the variation with respect to $\rho_r$ gives the condition
\begin{equation}\label{eom rho}
    \frac{\delta S_\text{EFT}}{\delta \rho_r} = 0 \ ,
\end{equation}
while, using $\rho_r = \dot \phi_r$, the variation with respect to $\phi_r$ is 
\begin{equation}\label{eq: variation action}
    -\int_{\mathcal M} \partial_t \left(\frac{\delta S_\text{EFT}}{\delta \rho_r} \right) \delta \phi_r + \int_{\partial \mathcal M} \frac{\delta S_\text{EFT}}{\delta \rho_r} \delta \phi_r \ ,
\end{equation}
where $\mathcal M$ is the spacetime manifold. Both of these terms must vanish separately. The bulk term only implies
\begin{equation}
    \frac{\delta S_\text{EFT}}{\delta \rho_r} = C(\textbf x) \ ,
\end{equation}
for some spatially dependent function $C(\textbf x)$. However, the boundary term in \eqref{eq: variation action} should also vanish. We cannot set $\delta\phi_r =0$ as a boundary condition as this is not compatible with the redundancy. Letting $\delta\phi_r$ vary freely at the boundary, the resulting variation term enforces $C(\textbf x) = 0$, recovering \eqref{eom rho} and demonstrating the equivalence of saddle points in the two descriptions.

\subsection{DKMS Transformations}\label{sec: KMS matter}

Having demonstrated a dictionary between covariants of the redundant and matter field parameterizations in Section~\ref{sec: dictionary}, it is instructive to discuss the realization of the DKMS transformation in the latter. By construction, the DKMS transformations map $G^R$ fibers into $G^R$ fibers. This implies that the DKMS transformations are uniquely defined on the local $G^R$ invariants $\phi_a$ and $\rho_r$, mapping them to functions of themselves.

Substituting \eqref{eq: relation between covariant derivatives in two approaches} into the DKMS transformations \eqref{eq: KMS transformations Goldstone approach ar basis} we obtain:
\begin{equation}\begin{aligned} \label{eq: KMS transformations Goldstone approach ar basis un-tilded}
   &e^{-i \phi_r'(x) \cdot T_r} \begin{pmatrix}  D_\mu \phi_r'(x) \cdot T \\  D_\mu \phi_a'(x) \cdot T \end{pmatrix} e^{i \phi_r'(x) \cdot T_r} \\
   &\qquad= -e^{-i \phi_r(y) \cdot T_r}\begin{pmatrix} \cosh\left( i \frac{\beta}{2}\mathcal{D}_t\right) & \frac{1}{2} \sinh\left( i \frac{\beta}{2}\mathcal{D}_t\right) \\ 2 \sinh\left( i \frac{\beta}{2}\mathcal{D}_t\right) & \cosh\left( i \frac{\beta}{2}\mathcal{D}_t \right) \end{pmatrix} \begin{pmatrix}  D_\mu \phi_r(y)  \cdot T \\ D_\mu \phi_a(y) \cdot T \end{pmatrix} e^{i \phi_r(y) \cdot T_r} \Bigg|_{y=-x} \ ,
\end{aligned}
\end{equation}
where we used the first equation in \eqref{eq: derivatives in tilde and un-tilde} in order to bring the exponentials $e^{\pm i \phi_r \cdot T_r}$ to the outside. Since the exponentials transform under DKMS as well, it would be incorrect to simply cancel them from each side of \eqref{eq: KMS transformations Goldstone approach ar basis un-tilded}. Nevertheless, by the argument above, the combination
\begin{equation}\label{exps}
    e^{i \phi_r'(x) \cdot T_r} e^{-i \phi_r(-x) \cdot T_r}
\end{equation}
must be invariant under $G^R$ and thus \eqref{eq: KMS transformations Goldstone approach ar basis un-tilded} uniquely defines the action of a DKMS transformation on $\rho_r$ and $D_\mu \phi_a$.
As a check we demonstrate the invariance of \eqref{exps} in Appendix~\ref{app: pertproof}.

As noted in Section~\ref{sec: dictionary}, the action in the redundant description is built out of $G^R$ adjoints, and the exponentials $e^{\pm i \phi_r \cdot T_r}$ will cancel. Therefore, at a practical level we may define new (we emphasize not DKMS) transformations stripped of the exponentials.

With the identification \eqref{identity}, we then read off
\begin{align}
	 \rho_r'(x)  &= -   \cosh\left(\frac{i \beta}{2} \mathcal{D}_t\right) \rho_r(y)  -  \, \frac{1}{2} \sinh\left(\frac{i \beta}{2} \mathcal{D}_t\right) D_0 \phi_a(y) \ \Bigg|_{y=-x} \ . \label{eq: first KMS transformation matter field approach}
\end{align}
At first sight, it is not obvious that the transformation of $D_\mu \phi_a$ will involve only $\rho_r$, due to the appearance of $D_i \phi_r$ on the right-hand side of \eqref{eq: KMS transformations Goldstone approach ar basis un-tilded}. However, using the identity \eqref{eq: relation commutator covariant derivative} we arrive at
\begin{align}\label{dkmsa}
	D_\mu \phi_a'(x) =  -   \cosh\left(\frac{i \beta}{2} \mathcal{D}_t \right) D_\mu \phi_a(y) - i \beta  F \left( \frac{i \beta}{2} \mathcal{D}_t \right) \left( \mathcal{F}_{0 \mu}(y) + \nabla_\mu \rho_r(y) \right) \ \Bigg|_{y=-x} \ ,
\end{align}
with $F(z) = \sinh(z) / z$. 

One may similarly find transformations for other ingredients, \textit{e.g.}, $\mathcal A_\mu$ via the transformation of $\nabla_\mu$.
Given the complete dictionary between the tilded and un-tilded covariant building blocks, any un-tilded effective action invariant under the new transformations \eqref{eq: first KMS transformation matter field approach} and \eqref{dkmsa} can be written in terms of a tilded action that is DKMS invariant by construction. 

In summary, we have demonstrated three results in this section. First, there exists a one-to-one correspondence between terms in the effective action written in the redundant parameterization and in the matter field approach. Second, we demonstrated the equivalence of the two approaches at the level of the path integral, in particular showing that the saddle points are equal despite the change of variables $\rho_r = D_0 \phi_r$ involving a time derivative. Finally, one may write down DKMS transformations for the fields of matter field approach. Next, we turn to a seemingly trivial, but conceptually important, phenomenon that is made manifest in the redundant variables.

\subsection{Non-dissipative Limit and Emergent Symmetry}\label{sec: Kelvin}

The time-independent redundancy \eqref{rsym} is of $r$-type, and if viewed as a symmetry of the effective action implies the conservation of an $a$-type current, whose physical meaning is not particularly transparent. On long distance scales however, the redundancy implies the emergence of a true symmetry. Concretely, consider the leading-order quadratic Lagrangian \eqref{eq: quadratic non Abelian eft} for distances $x$ and times $t$ such that 
\begin{equation}\label{dissip limit}
    \frac{x^2}{t} \gg \frac{\sigma}{\chi} \ ,
\end{equation}
that is, far away from the diffusive pole, where we can neglect the second term. We are then left with the non-dissipative Lagrangian, which is separable into terms depending only on $\phi_1$ and $\phi_2$, respectively.
\begin{equation}\label{dissip}
    \mathcal{L} = \chi \tilde D_t \phi_r \cdot \tilde D_t \phi_a = \mathcal{L}_1-\mathcal{L}_2 
    \, ,
\end{equation}
with $\mathcal{L}_i = \frac{\chi}{2} (\tilde D_t \phi_i)^2$. Invariance under $G^R$ thus implies an accidental, infinite-dimensional symmetry $G^R_1 \times G^R_2$, where $G_i^R$ acts as
\begin{equation}\label{asym}
    e^{i \phi_i(x) \cdot T}
  \longrightarrow 
    e^{i \phi_i(x) \cdot T} e^{i f(\textbf x)\cdot T}
    \ .
\end{equation}
These additional $a$-type transformations are associated with an $r$-type Noether current:
\begin{equation}\label{eq: current LO kelvin}
  J_r^\mu =   \chi \, \tilde D_t \phi_r \, \delta^\mu_0 \ .
\end{equation}
Because the spatial components of the current vanish, $J_r^0$ is independently conserved at each point in space. That is, there are an infinite number of conserved quantities. 

What happens in the matter-field approach? The redundancy (diagonal transformation) is not present, so that the appearance of the accidental symmetry is not trivial. However, having proven its equivalence with the redundant approach, the conservation law must remain. The Lagrangian is
\begin{equation}
     \mathcal{L} = \chi \, \rho_r \cdot D_t \phi_a\,,
\end{equation}
and the conservation of \eqref{eq: current LO kelvin} is simply the $\phi_a$ equation of motion, $\nabla_t \rho_r = 0$. The $\rho_r$ equation of motion along with the boundary conditions give $\phi_a = 0$ so that we finally get $\partial_t \rho_r = 0$. This is the trivial conservation of non-Abelian charge at each point in space when dissipation is neglected.\footnote{In general one has the term $\mathcal{L} \supset \chi \,  n(\rho_r) \cdot D_t \phi_a$, where the charge density $\chi\,  n(\rho_r)$ can be a non-linear function of the local chemical potential $\rho_r$. In this case we still have $\dot n = 0$.}

In the non-dissipative limit \eqref{dissip limit}, the separability of the Lagrangian \eqref{dissip} means that the system can be studied in the standard in-out (non-closed time path) formalism, where the Lagrangian is simply, say, $\mathcal{L}_1$. The Lagrangian is invariant under $G_1^R$, and the theory has a conserved current $J_1^\mu =   \chi \, \tilde D_t \phi_1 \, \delta^\mu_0$ associated with an infinite number of conserved quantities.
This system is off course completely trivial in the absence of diffusion---the charges remain stationary, and the charge distribution does not evolve. The situation becomes more interesting when energy and momentum transport are introduced, in addition to the transport of charge. This amounts to considering a charged fluid. 
With the complication that the fields $\phi_1$ become functions of an internal `fluid spacetime' rather than the physical spacetime \cite{Crossley:2015evo}, the symmetry \eqref{asym} becomes the so-called `chemical shift' symmetry, conserving the charge along the fluid flow \cite{Dubovsky:2011sj,Nicolis:2013lma}.
In other contexts, this symmetry has been used to attach the charges to any medium, for example a lattice \cite{Jain:2024ngx}, or to study pressureless dust in the universe \cite{Liang:2025gpp}. Because it restricts the mobility of charges in a medium, it is often called `fractonic.'

Perhaps the most insightful application of this class of symmetries comes when we consider the analogous redundancy that acts on the fluid coordinate fields themselves. The coset description of the transport of energy and momentum, arising from the conservation of spacetime symmetries, is beyond the scope of this work~\cite{futureus}.
However, in this case the non-dissipative limit remains non-trivial and is described by Euler's equation. The accidental symmetry and its implied conservation law is then Kelvin's theorem, ensuring conservation of vorticity along the flow \cite{Dubovsky:2005xd,Dubovsky:2011sj}.

\section{Discussion}

In this paper, we have defined a systematic procedure to construct Schwinger-Keldysh effective actions to describe the transport of non-Abelian internal charges. We compared two distinct but ultimately equivalent formulations: the redundant Goldstone mode parameterization and the matter field approach. By developing a precise dictionary between them and establishing their equivalence at the level of the path integral, we demonstrated that both approaches are consistent and interchangeable. 

A key result of our study is the implementation of the DKMS symmetry to all orders in $\hbar\omega/T$ both on the Goldstones and on the matter fields initiated in \cite{Akyuz:2023lsm}. As a consequence, we classified all invariant kernels compatible with unitarity and DKMS constraints. Our results provide a setup for systematically computing correlation functions and transport coefficients at arbitrary orders in $\hbar\omega/T$.

Our work also sheds light on the role of the redundancy in the Goldstone field approach to hydrodynamics. We showed that the same physical content can be reproduced without using redundant variables, raising doubts on the fundamental necessity of the redundancies. The matter field formulation not only avoids this complication, but also offers a conceptually cleaner description rooted in conventional EFT logic and the standard coset construction. This is achieved by effectively resumming a subset of non-linearities. For practical purposes however, the redundant approach still offers some benefits, as it is significantly simpler to use to define the KMS transformation and build the EFT.

These findings open up several directions for further exploration. On the one hand, completing the DKMS tensor calculus of Section~\ref{sec: kernels} is fundamental to systematically take into account higher derivative corrections. On the other, extending our approach to include energy-momentum transport will be essential for describing full relativistic hydrodynamics---see \textit{e.g.} Appendix A.2 in~\cite{Delacretaz:2023pxm} for a non-covariant attempt in this direction. Relatedly, it would be interesting to shed further light on the relation between the low-energy chemical symmetry and the standard Kelvin's theorem in non-dissipative fluid dynamics.

\acknowledgments

We are grateful to Can Onur Akyuz, Luca V. Delacretaz, Sergei Dubovsky, Garrett Goon, Greg Mathys, Ruchira Mishra, Alberto Nicolis,  Sergey Sibiryakov, and Stefan Stelzl for useful discussions. The work of RP is supported by the US Department of Energy grant DE-SC0010118. EF, AG, FN, and RR are partially supported by the Swiss National Science Foundation under contract 200020-213104 and through the National Center of Competence in Research SwissMAP.

\appendix
\section{Cubic kernels in the $(r,a)$ basis}\label{app:kern}

The three real cubic kernels \eqref{real_kern} in the $(r,a)$ basis are:
\begin{equation}\begin{aligned}
        \mathcal{R}^{(1)}_{\alpha_1\alpha_2\alpha_3} &= -\delta_{\alpha_1}^{r}\delta_{\alpha_2}^{r}\delta_{\alpha_3}^{a} \cosh\left[\frac{\beta \omega_1}{2}\right]\sinh\left[\frac{\beta \omega_2}{2}\right] +\delta_{\alpha_1}^{r}\delta_{\alpha_2}^{a}\delta_{\alpha_3}^{r} \sinh \left[\frac{\beta \omega_3}{2} \right]\\
        & + \delta_{\alpha_1}^{a}\delta_{\alpha_2}^{r}\delta_{\alpha_3}^{a} \frac{1}{2} \, \sinh\left[\frac{\beta \omega_1}{2}\right]\sinh\left[\frac{\beta \omega_2}{2}\right]-\delta_{\alpha_1}^{r}\delta_{\alpha_2}^{a}\delta_{\alpha_3}^{a} \frac{1}{2} \,\sinh\left[\frac{\beta \omega_1}{2}\right]\sinh\left[\frac{\beta \omega_2}{2}\right] \\
        &- \delta_{\alpha_1}^{a}\delta_{\alpha_2}^{a}\delta_{\alpha_3}^{a} \frac{1}{4} \, \sinh\left[\frac{\beta \omega_1}{2}\right]\cosh\left[\frac{\beta \omega_2}{2}\right]
\end{aligned}
\end{equation}
\begin{equation}\begin{aligned}
    \mathcal{R}^{(2)}_{\alpha_1\alpha_2\alpha_3} &= -\delta_{\alpha_1}^{r}\delta_{\alpha_2}^{r}\delta_{\alpha_3}^{a} \sinh\left[\frac{\beta \omega_1}{2}\right]\cosh\left[\frac{\beta \omega_2}{2}\right] +\delta_{\alpha_1}^{a}\delta_{\alpha_2}^{r}\delta_{\alpha_3}^{r} \sinh \left[\frac{\beta \omega_3}{2} \right]\\
        & - \delta_{\alpha_1}^{a}\delta_{\alpha_2}^{r}\delta_{\alpha_3}^{a} \frac{1}{2} \, \sinh\left[\frac{\beta \omega_1}{2}\right]\sinh\left[\frac{\beta \omega_2}{2}\right] +\delta_{\alpha_1}^{r}\delta_{\alpha_2}^{a}\delta_{\alpha_3}^{a} \frac{1}{2} \, \sinh\left[\frac{\beta \omega_1}{2}\right]\sinh\left[\frac{\beta \omega_2}{2}\right] \\
        &- \delta_{\alpha_1}^{a}\delta_{\alpha_2}^{a}\delta_{\alpha_3}^{a} \frac{1}{4} \, \cosh\left[\frac{\beta \omega_1}{2}\right]\sinh\left[\frac{\beta \omega_2}{2}\right]
\end{aligned}
\end{equation}
\begin{equation}\begin{aligned}
    \mathcal{R}^{(3)}_{\alpha_1\alpha_2\alpha_3} &= -\delta_{\alpha_1}^{r}\delta_{\alpha_2}^{r}\delta_{\alpha_3}^{a} 2 \, \sinh\left[\frac{\beta \omega_1}{2}\right]\sinh\left[\frac{\beta \omega_2}{2}\right] +\delta_{\alpha_1}^{a}\delta_{\alpha_2}^{a}\delta_{\alpha_3}^{r}  \sinh \left[\frac{\beta \omega_3}{2} \right]\\
        & + \delta_{\alpha_1}^{a}\delta_{\alpha_2}^{r}\delta_{\alpha_3}^{a}  \cosh\left[\frac{\beta \omega_1}{2}\right]\sinh\left[\frac{\beta \omega_2}{2}\right] +\delta_{\alpha_1}^{r}\delta_{\alpha_2}^{a}\delta_{\alpha_3}^{a}  \sinh\left[\frac{\beta \omega_1}{2}\right]\cosh\left[\frac{\beta \omega_2}{2}\right] \\
        &+ \delta_{\alpha_1}^{a}\delta_{\alpha_2}^{a}\delta_{\alpha_3}^{a} \frac{1}{4} \, \left(2 \sinh\left[\frac{\beta \omega_1}{2}\right]\sinh\left[\frac{\beta \omega_2}{2}\right]- 4 \cosh\left[\frac{\beta \omega_3}{2}\right]\right) 
\end{aligned}
\end{equation}
By \eqref{eq: kernel constraint 2} we see that $\mathcal{R}^{(1)}$ and $\mathcal{R}^{(2)}$ must come with a coefficient that is an odd function of $\vec \omega$, while $\mathcal{R}^{(3)}$ must come with an even coefficient.

The three imaginary cubic kernels \eqref{imag_kern} in the $(r,a)$ basis are:
\begin{equation}\begin{aligned}
        \mathcal{I}^{(1)}_{\alpha_1\alpha_2\alpha_3} &= -\delta_{\alpha_1}^{r}\delta_{\alpha_2}^{r}\delta_{\alpha_3}^{a} \cosh\left[\frac{\beta \omega_1}{2}\right]\sinh\left[\frac{\beta \omega_2}{2}\right] -\delta_{\alpha_1}^{r}\delta_{\alpha_2}^{a}\delta_{\alpha_3}^{r} \sinh \left[\frac{\beta \omega_3}{2} \right] \\
        & + \delta_{\alpha_1}^{a}\delta_{\alpha_2}^{r}\delta_{\alpha_3}^{a} \frac{1}{2} \, \sinh\left[\frac{\beta \omega_1}{2}\right]\sinh\left[\frac{\beta \omega_2}{2}\right] \\
        & -\delta_{\alpha_1}^{r}\delta_{\alpha_2}^{a}\delta_{\alpha_3}^{a}\left( \frac{1}{2} \, \sinh\left[\frac{\beta \omega_1}{2}\right]\sinh\left[\frac{\beta \omega_2}{2}\right]-  \cosh\left[\frac{\beta \omega_3}{2}\right]\right) \\
        &- \delta_{\alpha_1}^{a}\delta_{\alpha_2}^{a}\delta_{\alpha_3}^{a} \frac{1}{4} \, \sinh\left[\frac{\beta \omega_1}{2}\right]\cosh\left[\frac{\beta \omega_2}{2}\right]
\end{aligned}
\end{equation}
\begin{equation}\begin{aligned}
    \mathcal{I}^{(2)}_{\alpha_1\alpha_2\alpha_3} &= -\delta_{\alpha_1}^{r}\delta_{\alpha_2}^{r}\delta_{\alpha_3}^{a} \sinh\left[\frac{\beta \omega_1}{2}\right]\cosh\left[\frac{\beta \omega_2}{2}\right] -\delta_{\alpha_1}^{a}\delta_{\alpha_2}^{r}\delta_{\alpha_3}^{r} \sinh \left[\frac{\beta \omega_3}{2} \right]\\
        & - \delta_{\alpha_1}^{a}\delta_{\alpha_2}^{r}\delta_{\alpha_3}^{a} \left( \frac{1}{2} \, \sinh\left[\frac{\beta \omega_1}{2}\right]\sinh\left[\frac{\beta \omega_2}{2}\right] - \cosh\left[\frac{\beta \omega_3}{2}\right] \right) \\
        & + \delta_{\alpha_1}^{r}\delta_{\alpha_2}^{a}\delta_{\alpha_3}^{a} \frac{1}{2} \, \sinh\left[\frac{\beta \omega_1}{2}\right]\sinh\left[\frac{\beta \omega_2}{2}\right] \\
        &- \delta_{\alpha_1}^{a}\delta_{\alpha_2}^{a}\delta_{\alpha_3}^{a} \frac{1}{4} \, \cosh\left[\frac{\beta \omega_1}{2}\right]\sinh\left[\frac{\beta \omega_2}{2}\right]
\end{aligned}
\end{equation}
\begin{equation}\begin{aligned}
    \mathcal{I}^{(3)}_{\alpha_1\alpha_2\alpha_3} &= -\delta_{\alpha_1}^{r}\delta_{\alpha_2}^{r}\delta_{\alpha_3}^{a} 2 \sinh\left[\frac{\beta \omega_1}{2}\right]\sinh\left[\frac{\beta \omega_2}{2}\right] -\delta_{\alpha_1}^{a}\delta_{\alpha_2}^{a}\delta_{\alpha_3}^{r}  \sinh \left[\frac{\beta \omega_3}{2} \right]\\
        & + \delta_{\alpha_1}^{a}\delta_{\alpha_2}^{r}\delta_{\alpha_3}^{a}  \cosh\left[\frac{\beta \omega_1}{2}\right]\sinh\left[\frac{\beta \omega_2}{2}\right] +\delta_{\alpha_1}^{r}\delta_{\alpha_2}^{a}\delta_{\alpha_3}^{a}  \sinh\left[\frac{\beta \omega_1}{2}\right]\cosh\left[\frac{\beta \omega_2}{2}\right] \\
        &+ \delta_{\alpha_1}^{a}\delta_{\alpha_2}^{a}\delta_{\alpha_3}^{a} \frac{1}{2} \, \sinh\left[\frac{\beta \omega_1}{2}\right]\sinh\left[\frac{\beta \omega_2}{2}\right]
\end{aligned}
\end{equation}
By \eqref{eq: kernel constraint 2}, $\mathcal{I}^{(1)}$ and $\mathcal{I}^{(2)}$ must come with an even coefficient, while $\mathcal{I}^{(3)}$ must come with an odd coefficient.


\section{Invariance of the Path Integral Contour under DKMS}\label{app: KMS contour}

In this appendix, we briefly discuss a subtlety when implementing the dynamical KMS symmetry in the path integral construction of our effective field theory. 
Indeed, in Section~\ref{sec: kernels} we built DKMS invariant actions, but did not worry about the invariance of the integration contour. 

This analysis is easier to perform in frequency space, where the real scalar fields $\phi_i(t)$, $i=1,2$, are mapped to complex fields $\hat\phi_i(\omega)$ that respect the reality condition $\hat\phi_i^*(\omega) =\hat\phi_i(-\omega) $. In the rest of this appendix, all fields are in frequency space, and we thus drop the hat notation.

We can then treat $\bar \phi_i(\omega) \equiv \phi_i(-\omega)$ and $\phi_i(\omega)$ as independent complex variables and write the path integral as an integration on a domain $\Gamma$ in the complex plane 
\begin{equation}
   \int_\Gamma D\phi_1 D\bar\phi_1 D\phi_2 D\bar\phi_2 \, e^{i S[\phi_1,\bar\phi_1,\phi_2,\bar\phi_2]} \ ,
\end{equation}
where the integration domain $\Gamma$ is defined by the condition $\phi_i^*(\omega)=\bar\phi_i(\omega)$, for $i=1,2$, and we only integrate over fields with $\omega > 0$. Importantly, $S$ depends only on $\phi_i$ and $\bar \phi_i$, not on their complex conjugates, and is therefore a holomorphic function.

The DKMS transformations read:
\begin{equation}\label{eq:KMS fourier}
    \begin{pmatrix}
\phi_1'(\omega) \\
\phi_2'(\omega)
\end{pmatrix}
=
\begin{pmatrix}
e^{-\beta \omega/2} & 0 \\
0 & e^{\beta \omega/2}
\end{pmatrix}
\begin{pmatrix}
\bar\phi_1(\omega) \\
\bar\phi_2(\omega)
\end{pmatrix}, \quad \begin{pmatrix}
\bar \phi_1'(\omega) \\
\bar \phi_2'(\omega)
\end{pmatrix}
=
\begin{pmatrix}
e^{\beta \omega/2} & 0 \\
0 & e^{-\beta \omega/2}
\end{pmatrix}
\begin{pmatrix}
\phi_1(\omega) \\
\phi_2(\omega)
\end{pmatrix}\, .
\end{equation}
Here we are not representing spatial coordinates $\bold{x}$ as they will not play an important role in the discussion. 
In the case of Goldstone fields, the measure of integration is the Haar measure. This measure is invariant under translations and thus is invariant under DKMS. 

However, these transformations do not leave the domain of integration invariant as the condition $\phi^*_i(\omega)=\bar\phi_i(\omega)$ is not preserved by  \eqref{eq:KMS fourier}. The new domain of integration $\Gamma'$ is defined by the conditions
\begin{equation}\label{condss}
    e^{-\beta \omega/2} \bar\phi_1(\omega)=e^{\beta \omega/2} \phi^*_1(\omega) \ , \quad e^{\beta \omega/2} \bar\phi_2(\omega)=e^{-\beta \omega/2} \phi^*_2(\omega) \ .
\end{equation} 
As \eqref{eq:KMS fourier} only mixes the modes 
of the same frequency,
we can restrict our analysis to the integration over the modes associated to 
some fixed value of $\omega$. 

We therefore have integrals of the form:
\begin{equation}\label{eq: integral old contour omega}
    \int_\gamma  d\phi_1(\omega)\wedge d\bar\phi_1(\omega)\wedge d\phi_2(\omega)\wedge d\bar\phi_2(\omega) \, \,F[\phi_1,\bar\phi_1,\phi_2,\bar\phi_2]\, ,
\end{equation}
with $F$ an holomorphic function, and the domain of integration $\gamma$ is again defined by the constraint $\phi_i^*(\omega)=\bar\phi_i(\omega)$. Notice in particular that by the holomorphy of F,
\begin{equation} \label{eq: integrant closed form}
    d\left( d\phi_1(\omega)\wedge d\bar\phi_1(\omega)\wedge d\phi_2(\omega)\wedge d\bar\phi_2(\omega) \, \,F[\phi_1,\bar\phi_1,\phi_2,\bar\phi_2] \right) =0 \, .
\end{equation}
By Stokes' theorem, \eqref{eq: integrant closed form} implies that when the integrand is integrated over a closed contour, it vanishes (as long as it is analytic on the interior, which is guaranteed by analyticity on the original contour and in the limit $\omega \ll T$). This enables us to relate the integral \eqref{eq: integral old contour omega} on the domain $\gamma$ to the same integral on the domain $\gamma'$, defined by the conditions \eqref{condss}:
\begin{equation}
    \begin{split}
        \int_\gamma  d\phi_1(\omega)\wedge d\bar\phi_1(\omega)&\wedge d\phi_2(\omega)\wedge d\bar\phi_2(\omega) \, \,F[\phi_1,\bar\phi_1,\phi_2,\bar\phi_2] \\ &= \int_{\gamma'}  d\phi_1(\omega)\wedge d\bar\phi_1(\omega)\wedge d\phi_2(\omega)\wedge d\bar\phi_2(\omega) \, \,F[\phi_1,\bar\phi_1,\phi_2,\bar\phi_2] \\
        &+ \textnormal{boundary terms.}
    \end{split}
\end{equation}

The boundary terms come from the behaviour of the integrand at the large-field boundary. They are the result of integrating $F$ over the boundary at infinity of the interpolating family between 
$\gamma$ and $\gamma'$. More explicitly, this boundary is parametrized as $\{|\bar\phi_1(\omega)|^2 +|\bar\phi_2(\omega)|^2=R^2, \, \phi_1^*(\omega) = e^{s \beta \omega} \bar\phi_1(\omega), \, \phi_2^*(\omega) = e^{-s \beta \omega} \bar\phi_2(\omega)\,|\, s \in [0,1],\, R \to \infty\}$. Assuming $F$ decays fast enough with $R$, we can drop these terms and we obtain that the integral over $\gamma$ and $\gamma'$ coincide.

Performing the same argument for all frequencies $\omega$, we have therefore proved that although the transformation maps the path integral to a different contour, one can map it back to the original contour using Stokes' theorem. This is true assuming the integrant decays fast enough at infinity, which we generically expect to be verified for an action associated to a healthy EFT. Therefore, the DKMS transformation is a symmetry of the path integral.


\section{Equivalence of the equations of motion in the non-Abelian case}\label{app: non Abelian EOM}

We have proved in Section \ref{sec: EOM equivalence} that in the Abelian case, the equations of motions in the redundant and matter field approaches have the same solutions once we take into account the boundary conditions. The saddle points are therefore equivalent. In this appendix, we generalize the proof to the non-Abelian case.

In the non-Abelian case, $\rho_r$ is no longer linear in $\phi_r$. Throughout this appendix we use a condensed notation where all fields are implicitly contracted with the $T_r$ (\textit{e.g.} $\phi_r = \phi_r^I T_r^I$). Using $\rho_r= i e^{i \phi_r} \partial_{\mu} e^{-i \phi_r } + \mathcal{A}_{t}$ and calling $\text{EOM}_{\rho_r} \equiv \frac{\delta S}{\delta \rho_r}$, one can expand $S[\phi_r + \delta \phi_r]$ to linear order in $\delta\phi_r$ to obtain
\begin{equation}\label{eq: S non Abelian variation}
    \begin{split}
        \delta S=- \int d^4 x \int_0^1 d\alpha \, \Bigg[ &e^{i \alpha \phi_r} \delta \phi_r e^{i(1-\alpha) \phi_r}\partial_t \left(e^{-i \phi_r} \right) \text{EOM}_{\rho_r} \\ &- e^{i \phi_r}  \partial_t  \left( e^{-i \alpha \phi_r} \delta\phi_r e^{- i (1-\alpha)  \phi_r } \right) \text{EOM}_{\rho_r} \Bigg]\, ,
    \end{split}
\end{equation}
where we used
\begin{equation} \label{eq: alpha}
    \delta(e^{i \phi_r\cdot T_r})=i \int_0^1 d\alpha \, e^{ i (1-\alpha)\phi_r}\, \delta \phi_r\, e^{ i \alpha \phi_r} \,.
\end{equation}
Integrating \eqref{eq: S non Abelian variation} by parts, we obtain the bulk equation for $\phi_r$:
\begin{equation}\label{eq: EOM non Abelian 1}
    \int_0^1 d\alpha \,\Bigg[ e^{i(1-\alpha) \phi_r}\partial_t \left(e^{-i \phi_r} \right) \text{EOM}_{\rho_r} e^{i \alpha \phi_r} + e^{- i (1-\alpha)  \phi_r } \partial_t \left( \text{EOM}_{\rho_r} e^{i \phi_r} \right)e^{-i \alpha \phi_r} \Bigg] =0\, ,
\end{equation}
as well as the constraint from the $t=+\infty$ limit:
\begin{equation}\label{eq: boundary constraint nonAbelian}
   \left.  \int_0^1 d\alpha \, e^{ i \alpha \phi_r} \left( e^{i \phi_r} \text{EOM}_{\rho_r}  e^{-i \phi_r }   \right)  e^{ -i \alpha\phi_r}\right|_{t=+\infty} =0\, .
\end{equation}
Performing some simple mathematical manipulation, we can rewrite \eqref{eq: EOM non Abelian 1} as
\begin{equation}
    \int_0^1 d\alpha \, e^{ -i  \alpha \phi_r} \Bigg( \nabla_t \text{EOM}_{\rho_r} -i \left[ i\, e^{i \phi_r} \partial_{0} e^{-i \phi_r} + \mathcal{A}_{0}, \text{EOM}_{\rho_r}  \right]  \Bigg) e^{ i \alpha \phi_r}=0 \, .
\end{equation}
The quantity in parentheses is $\mathcal{D}_t \text{EOM}_{\rho_r}$ so that we may use Eq. \eqref{eq: derivatives in tilde and un-tilde} and the change of variable $\alpha \to 1- \alpha$ to get
\begin{equation}
    \int_0^1 d\alpha \, e^{ i  \alpha \phi_r} \partial_t \left( e^{i \phi_r\cdot T_r} \text{EOM}_{\rho_r}  e^{-i \phi_r \cdot T_r}   \right) e^{ -i \alpha \phi_r} =0 \, .
\end{equation}
This has the interpretation of the quantity in the parentheses being integrated over a partial orbit under the action of $\phi_r$, which can be shown to only vanish if the orbit is closed (or if the quantity itself vanishes). For any $\phi_r$ one can choose local coordinates on  $G_r$ such that the orbit is not closed. Thus we generically have
\begin{equation}
    \partial_t \left( e^{i \phi_r} \text{EOM}_{\rho_r}  e^{-i \phi_r}   \right)=0 \implies \text{EOM}_{\rho_r} = e^{-i \phi_r} C(\textbf x) e^{i \phi_r} \, ,
\end{equation}
with $C(\textbf x)$ now in the adjoint. Using the constraint  \ref{eq: boundary constraint nonAbelian}, that forces $\text{EOM}_{\rho_r}$  to vanish at $t \to \infty$, the right-hand side must vanish identically, and the proof is complete.

\section{KMS transformations to leading order}\label{app: pertproof}
In this section we will check to first order in derivatives and quadratic order in the fields the DKMS transformations of  $\phi_r$ and $\phi_a$. We will show that the latter, as well as the combination (\ref{exps}), is invariant under the local right action (\ref{gaugeg}). To this end we start from the two different parameterizations of the coset,
\begin{equation}\tag{\ref{cosetforreal}}
    \tilde \Omega = e^{i\phi_a\cdot T_a}e^{i\phi_r\cdot T_r}=e^{i\phi_1 \cdot T_1}e^{i\phi_2\cdot T_2} \ ,
\end{equation}
with the generators 
\begin{equation}\tag{\ref{gens}}
T_r = T_1 + T_2 \ , \qquad T_a = \frac{1}{2} \left(T_1 - T_2 \right) \ .
\end{equation}
Under the action of $G^R$, the $\phi_a$ field is invariant while,
\begin{equation}\label{phirGR}
    \phi_r^h=\phi_r-f-\frac{i}{2} [\phi_r,f]+\cdots \ ,
\end{equation}
to quadratic order in $\phi_r$ and the time-independent $f(\textbf x)$.
The relations between the fields to quadratic order are,
\begin{equation}
\begin{split}
    &\phi_r = \frac{1}{2}(\phi_1+\phi_2)+\mathcal{O}(\phi^3),\\
    &\phi_a=\phi_1-\phi_2-\frac{i}{2}[\phi_1,\phi_2]+\mathcal{O}(\phi^3) \ .
\end{split}
\end{equation}
Throughout this appendix we use a condensed notation where all fields are implicitly contracted with generators of $G$ ($\phi = \phi^I T^I$). 
The DKMS transformations for the $1$- and $2$-fields are,\footnote{The dot represents the derivative with respect to the $0-$th component of the argument which is $-t$.}
\begin{equation}
    \begin{split}
        &\phi_1'(x)=\phi_1(-t+i\beta/2, - \textbf x)=\phi_1(-x)+\frac{i}{2}\beta \,\dot{\phi_1}(-x)+\mathcal{O}((\beta\partial_t)^2) \ ,\\ &\phi_2'(x)=\phi_2(-t-i\beta/2, - \textbf x)=\phi_2(-x)-\frac{i}{2}\beta \,\dot{\phi_2}(-x)+\mathcal{O}((\beta\partial_t)^2) \ .
    \end{split}
\end{equation}
Using the inverse relations,
\begin{equation}
\begin{split}
    &\phi_1 = \phi_r + \frac{\phi_a}{2}-\frac{i}{4}[\phi_r,\phi_a]+\mathcal{O}(\phi^3) \ ,\\
    &\phi_2 = \phi_r - \frac{\phi_a}{2}+\frac{i}{4}[\phi_r,\phi_a]+\mathcal{O}(\phi^3) \ ,
\end{split}
\end{equation}
these induce the following transformations on the $r$- and $a$-fields:
\begin{equation}
\begin{split}
    &\phi_r'(x)=\phi_r(-x)+i\frac{\beta}{4}\dot{\phi}_a(-x)+\frac{\beta}{8} [\dot{\phi_r},\phi_a](-x)+\frac{\beta}{8} [\phi_r,\dot{\phi}_a](-x)+\mathcal{O}((\beta \partial_t)^2,\phi^3) \ ,\\
    &\phi_a'(x)=\phi_a(-x)+i\beta\dot{\phi}_r(-x)-\frac{\beta}{2} [\phi_r,\dot{\phi_r}](-x)+\frac{\beta}{8} [\phi_a,\dot{\phi}_a](-x)+\mathcal{O}((\beta \partial_t)^2,\phi^3) \ .
\end{split}
\end{equation}
One can check explicitly the invariance of $\phi_a'$ under (\ref{phirGR}). Finally, we check the invariance of (\ref{exps}). This immediately follows from the rewriting,
\begin{equation}
    e^{i \phi_r'(x)} e^{-i \phi_r(-x)}=\exp\bigg(-\frac{\beta}{4}\dot{\phi}_a(-x)+\frac{i}{8}\beta [\dot{\phi}_r,\phi_a]+\mathcal{O}((\beta\partial_t)^2,\phi^3)\bigg) \ .
\end{equation}




\bibliographystyle{JHEP}
\bibliography{biblio}

\end{document}